\documentclass[preprint,12pt]{elsarticle}

\usepackage{amssymb}

\usepackage{xcolor}
\AtBeginDocument{\pagenumbering{arabic}}
\usepackage{stackrel}
\usepackage{amssymb}
\usepackage{amsmath}
\usepackage{multirow}
\usepackage{mathtools}

\usepackage[utf8]{inputenc}
\usepackage[T1]{fontenc}

\usepackage{nomencl}
\usepackage{etoolbox}
\makenomenclature
\renewcommand\nomgroup[1]{%
  \item[\bfseries
  \ifstrequal{#1}{G}{Greek Symbols}{
  \ifstrequal{#1}{S}{Subscripts}{
  \ifstrequal{#1}{P}{Superscripts}{}}}
]}

\newlength{\margin}
\begin{document}

\begin{frontmatter}



\title{Fluid dynamics of mixing in the tanks of small vanadium redox flow batteries: Insights from order-of-magnitude estimates and transient two-dimensional simulations}


\author[aff1]{Pablo A. Prieto-D\'iaz}
\ead{pabloangel.prieto@uc3m.es}
\author[aff2]{Santiago E. Ib\'a\~{n}ez}
\ead{santiagoenrique.ibanez@repsol.com}
\author[aff1]{Marcos Vera}
\ead{marcos.vera@uc3m.es}
\affiliation[aff1]{organization={Departamento de Ingenier\'{\i}a T\'{e}rmica y de Fluidos, Universidad Carlos III de Madrid},
            addressline={Avd. de la Universidad 30},
            city={Legan\'{e}s},
            postcode={28911},
            state={Madrid},
            country={Spain}}
\affiliation[aff2]{organization={Repsol Technology Lab},
            addressline={Paseo de Extremadura, km 18},
            city={M\'{o}stoles},
            postcode={28939},
            state={Madrid},
            country={Spain}}

\begin{abstract}
This paper investigates the fluid dynamics of mixing in the tanks of small-scale vanadium redox flow batteries. These systems use two redox pairs dissolved in separate electrolytes to convert electrical energy into chemical energy, a process that can be reversed in an efficient way to restore the initial electrical energy with little or negligible chemical losses. After flowing through the electrochemical cell, the electrolytes are stored in separate tanks, where they discharge as submerged jets with small temperature and composition changes compared to the electrolyte already present in the tanks. The subsequent mixing process is critical for battery performance, as imperfect mixing tends to reduce the energy capacity and may lead to asymmetric battery operation. The analysis starts using order-of-magnitude estimates to determine the conditions under which the mixing process is dominated by momentum or buoyancy. Transient two-dimensional simulations are used to illustrate the different flow regimes that emerge in the tanks under laminar flow conditions. The results show that, contrary to the common assumption, the electrolytes do not mix well in the tanks. In the presence of high-momentum---and, specially, positively buoyant---jets, a significant fraction of the electrolyte remains unmixed and unreacted for long periods, thus reducing the energy capacity. The results also show that the availability of reliable electrolyte properties is crucial for the accuracy of the numerical simulations, as, under the mixed convection conditions that typically prevail in vanadium redox flow batteries, small density variations can significantly impact the long-term mixing of the electrolytes. In particular, in momentum-dominated flows the cumulative effect of density changes over time eventually leads to flow instabilities that significantly promote mixing; therefore, they should be taken into account in future studies to optimize tank design.

\end{abstract}

\begin{graphicalabstract}
\includegraphics{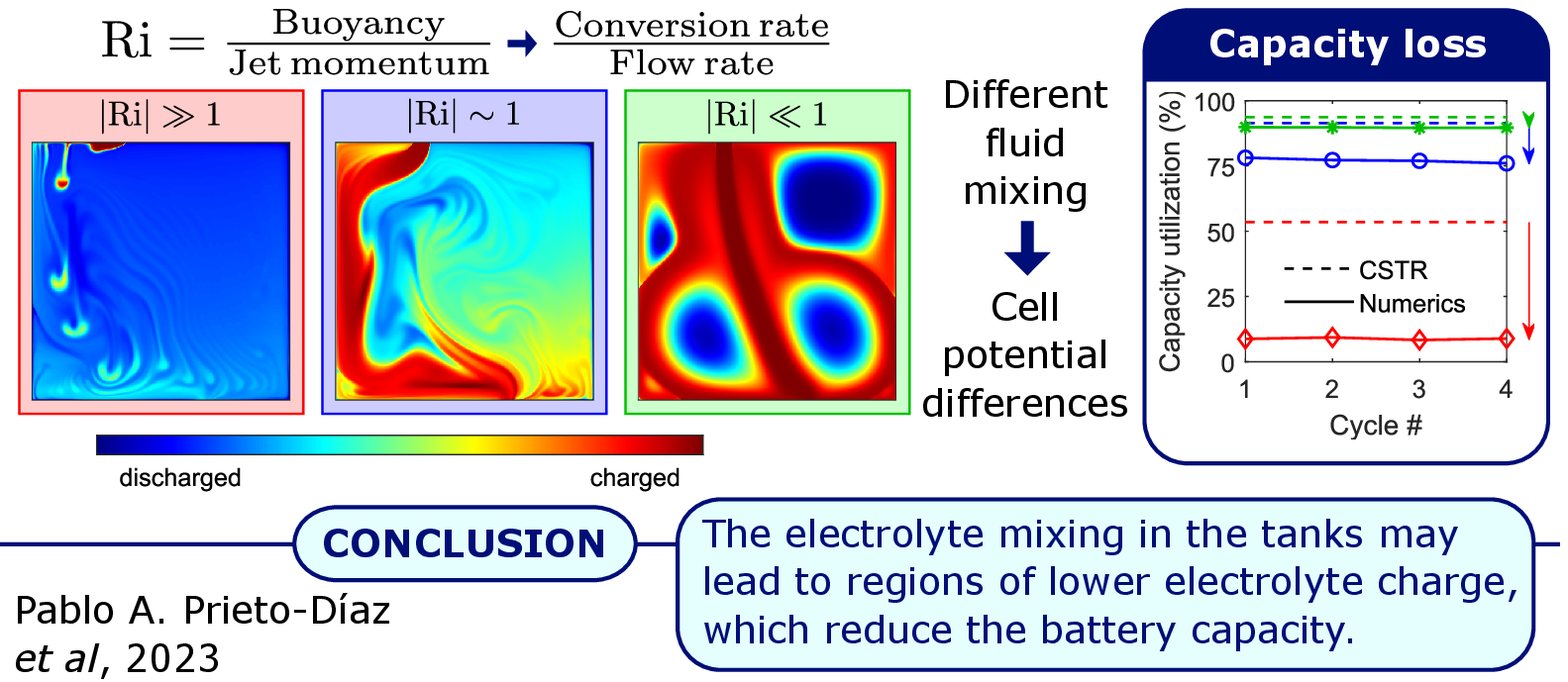}
\end{graphicalabstract}

\begin{highlights}
\item A new model is proposed to investigate mixing in the tanks of vanadium redox flow batteries.
\item Numerical integrations are performed with the finite element method for 2D laminar flows.
\item Order of magnitude estimates are used to unveil the dominant physical mechanisms.
\item Buoyancy affects mixing strongly buy asymmetrically in both tanks, particularly for low flow and high conversion rates.
\item Having reliable electrolyte data is crucial for the accuracy of the numerical results.
\end{highlights}


\end{frontmatter}



\section{Introduction}\label{section_intro}

\begin{figure}[t!]
	\centering
	\includegraphics{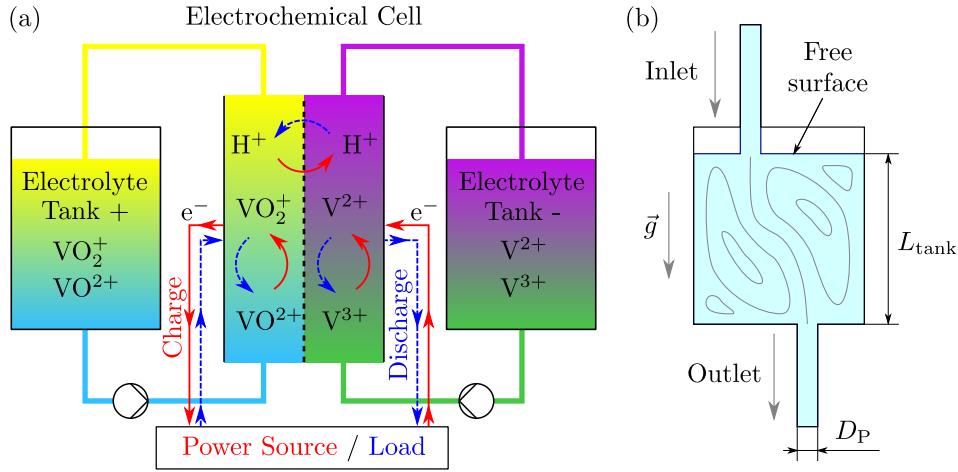}
	\caption{Schematic representation of (a) a vanadium redox flow battery and (b) detailed view of the two-dimensional tank geometry and dimensions used in this study.}
	\label{figure_scheme_tank}
\end{figure}

The use of renewable energy has experienced a rapid growth in the last decades, primarily aimed at mitigating the carbon emissions associated with electricity generation from conventional fossil fuels \cite{Xu2015}. Although promising, these technologies remain based on intermittent and variable energy sources, such as wind or solar, making it necessary to develop innovative and more efficient energy storage systems to ensure the stability of the grid \cite{Skyllas2011}. Often included among the most promising technologies, Redox Flow Batteries (RFBs) stand out for their flexible modular design and operation, characterized by the independence between power and energy storage capacities. The former is related to the size of the electrochemical conversion cells and stacks, whereas the latter depends on the volume of the electrolyte storage tanks. Among the various chemistries used in redox flow batteries, all-Vanadium system (VRFBs) proposed by Skyllas-Kazacos et~al.~\cite{Skyllas1986} has gained commercial viability owing to its impact of cross-mixing, high energy efficiency and long life cycle \cite{Skyllas2011, Skyllas2016}. VRFBs employ two redox couples---$\rm V^{2+}/V^{3+}$ as the anolyte and $\rm VO^{2+}/VO_2^+$ as the catholyte---dissolved in highly concentrated aqueous sulphuric acid solutions that are stored in two separated tanks. A schematic representation of a VRFB with indication of the two-dimensional tank geometry and dimensions used in this study is shown in Figure~\ref{figure_scheme_tank}.

Mathematical modeling is a powerful tool to expedite the development of new generation VRFBs. Several models have been proposed to analyze the performance of these systems and improve their design, mostly focused on the electrochemical cells or stacks \cite{Zheng2014}. One of the first models was proposed by Li and Hikihara \cite{Li2008}, who derived a zero dimensional dynamical model to simulate the electrochemical and hydraulic performance of VRFBs. Shortly after, Shah et~al.~\cite{Shah2008} developed a new model retaining species concentrations and current density gradients within the cell, followed by You et~al.~\cite{You2009}, who enhanced and applied this model to predict the effect of state of charge (SoC) variations in the system. Al-Fetlawi et~al.~\cite{Al-Fetlawi2009} incorporated temperature variations in a two-dimensional model, and investigated with Shah et~al.~\cite{Al-Fetlawi2010, Shah2010} both hydrogen and oxygen evolution. Vynnycky~\cite{Vynnycky2011} proposed an asymptotic model, while Ma et~al.~\cite{Ma2011} used a three-dimensional model of the negative electrode to determine the effect of the velocity field on the species concentrations, overpotential and transfer current density distributions. You et~al.~\cite{You2011} and Tang et~al.~\cite{Tang2011} modelled ion diffusion through the membrane, and Knehr et~al.~\cite{Knehr2012} incorporated convection and migration effects. Tang et~al.~\cite{Tang2012-2} used a non-isothermal zero-dimensional model to optimize the operation temperature, a second model that included heating due to the self-discharge reactions \cite{Tang2012}, and a third one to investigate pump losses and the effect of flow rate on VRFBs \cite{Tang2014}. A similar approach was followed by Xiong et~al.~\cite{Xiong2013}. After these pioneering works, further models have been developed addressing ion crossover \cite{Zhou2015,Hao2019,Pugach2018}, thermal management to improve design and operation conditions \cite{We2014,Wei2014,Trovo2019-2,Trovo2019}, three-dimensional geometries \cite{Oh2015}, flow rate optimization \cite{Khazaeli2015,Kim2018}, industrial scale systems \cite{Konig2018}, asymmetric electrode design \cite{Lu2019}, membraneless VRFBs \cite{Ibanez2021}, to the most recent advanced multi-physics models (see, e.g., \cite{Munoz2022,Munoz2023} for two recent contributions from our group). Comprehensive discussions of the most relevant works on RFB modeling can be found in the various reviews available in the literature \cite{Xu2015, Weber2011, Esan2020}.

Unlike the strong attention paid to the electrochemical cells, with the development of models of increasing complexity over the last decade, the flow of the electrolytes in the tanks has received much less attention in the literature. Research in this area has generally focused on experimental work addressing industrial strategies for rebalancing and mixing both tanks \cite{Wang2017,Bhattarai2018}, four-tank configurations \cite{Liu2019}, or gravity-induced flows \cite{Chen2016}. Regarding the modeling attempts, the typical approach has been to assume perfect mixing in the tanks throughout the charge-discharge cycle (the so-called Continuous Stirred Tank Reactor (CSTR) models), i.e., that the electrolyte immediately mixes with the entire liquid volume upon entering the tank. More recently, Nemani et~al.~\cite{Nemani2017} stood out the importance of the mixing process in the tanks, while Wang et~al.~\cite{Wang2019} investigated the fluid dynamics inside the tanks, highlighting the importance of using baffles. But they ignored the effect of buoyancy and considered very low Reynolds numbers.

The physicochemical properties of vanadium electrolytes change over time during the operation of the battery, mainly due to changes in the SoC. The properties of vanadium electrolytes, such as density \cite{Mousa2003,Rahman2009,Xu2014-2,Skyllas2016}, viscosity \cite{Mousa2003,Kausar2002,Skyllas2016}, and specific heat \cite{Qin2010}, have been reported as a function of species concentration and/or temperature. However, reports focusing on the dependence of these properties on the SoC are scarce. Changes in viscosity play a crucial role in the flow cell, as they alter the pressure drop across the porous electrodes and are related to the diffusivity of the active species. Consequently, the viscosity of vanadium electrolytes has been thoroughly studied in the literature \cite{Xu2014, Li2018}. By contrast, density variations have been largely overlooked. Skyllas et~al.~\cite{Skyllas2016} reported density data of 2 M vanadium solutions in 5 M sulfuric acid as a function of SoC and temperature, but without describing the experimental procedures or providing error estimations, which are typically large in this type of measurements. The negative electrolyte data reported by Ressel et~al.~\cite{Ressel2018} were consistent with those measurements. According to these studies, variations in density with SoC could reach up to 2\%. These small changes, as insignificant they might seem for the flow in the cells, could be crucial in the tanks, where mixed convection prevails with buoyancy forces that can be dominant for sufficiently low velocities and high conversion rates.

Buoyancy effects appear in all variable-density flows subject to gravity. The importance of buoyancy grows with density changes and may be characterized by the so-called Richardson number
\begin{equation}
\textrm{Ri} = \frac{\Delta \rho}{\rho} \frac{gh}{u^2} \label{eq_Ri_intro}
\end{equation}
where $\Delta \rho$ denotes the density difference over a characteristic (often vertical) length scale, $h$, in a flow with characteristic velocity $u$ \cite{Cebeci88}. In a mixed convection problem, such as the one considered here, the Richardson number represents the importance of buoyancy (i.e., natural convection) relative to forced convection (i.e., flow momentum). Thus, as $\textrm{Ri} \rightarrow 0$, so does the importance of buoyancy. The above expression implies that significant buoyancy effects can occur either for sufficiently large density changes $\Delta \rho/\rho$ or significant values of $gh/u^2$. Therefore, in the electrolyte tanks of VRFBs, small density variations, of order $\Delta \rho/\rho \sim 0.02$, may give rise to significant buoyancy effects for sufficiently large tanks, $h \sim 1~\textrm{m}$, and moderate flow velocities, $u \sim 1~\textrm{m}/\textrm{s}$.

Buoyant flows in the tanks of VRFBs involve a range of fluid-mechanical processes such as submerged buoyant jets, stratified flows, baroclinic vorticity production, flow instabilities, mixed-convection boundary layers, and more \cite{Jaluria1980}. These phenomena can have a significant impact on the mixing of electrolytes in the tanks, thereby affecting the overall performance of the battery. The aim of this study is to develop a comprehensive mathematical model that can be used to numerically simulate the mixed-convection flow of vanadium electrolytes in VRFB tanks and analyze its effect on system performance. Additionally, the investigation of the model's sensitivity to variations in the electrolyte properties highlights the need for a more precise experimental characterization of these properties.

The paper is structured as follows. The mathematical model is presented in Section \ref{section_model}, including the description of the tanks as two-dimensional reservoirs and of the electrochemical cell using a zero-dimensional model. Section \ref{section_dimensional_analysis} provides order of magnitude estimates that anticipate the type of mixed-convection flow as a function of the operating conditions, tank geometry, and electrolyte properties. The numerical method, based in the Finite Element Method (FEM), is presented in Section \ref{section_numerical_method}. Numerical results illustrating the fluid dynamics and mixing of the electrolytes and their implications in cell operation are reported in Section \ref{section_results} for various case studies. Finally, the conclusions are presented in Section \ref{section_conclusions}.

\section{Mathematical model}\label{section_model}

In this section, we present the mathematical model, which is divided into two main elements: the electrolyte tanks and the electrochemical cell. The flow of the electrolytes in the tanks is described by the Navier-Stokes equations, supplemented by the species mass conservation equations and the energy equation, written here in terms of temperature by assuming a constant specific heat. In contrast, the cell is described by a zero-dimensional model that accounts for the main electrochemical reactions and provides the corresponding changes in electrolyte composition (SoC) and temperature in a single cell pass. The cell model is coupled to the positive and negative tanks through the inlet and outlet boundary conditions.

The tanks could be modelled either as two- or three-dimensional reservoirs. However, the results presented below presume a two-dimensional (2D) geometry due to the strong computational cost associated with three-dimensional (3D) simulations. This simplification yields solutions that may differ quantitatively from the behavior of real 3D systems, but whose overall flow behaviour, transient buoyancy effects, and composition evolution are expected to be qualitatively correct. Figure \ref{figure_scheme_tank}b shows a schematic view of the tank geometry considered in this study. The tanks have a common square-shaped fixed geometry and are supposed to have the inlet located at the top and the outlet at the bottom, placed respectively at 1/3 and 2/3 of the tank width, $L_\textrm{tank}$, to improve mixing. Moreover, the volumetric flow rates of both electrolytes are assumed identical and fixed over time.

To further simplify the analysis, the flow is assumed to remain laminar and unsteady, with jet Reynolds numbers in the range $\textrm{Re}_\textrm{in}=\rho u_\textrm{in} D_\textrm{P}/\mu \in [5,188]$ in all cases under study; based on the fluid density, $\rho$, the average velocity of the discharging jet, $u_{\textrm{in}}$, the pipe diameter, $D_{\textrm{P}}$, and the fluid viscosity, $\mu$. This choice allows focusing on the mixed convection flow associated with the changes in electrolyte properties, avoiding the complexities entailed by turbulent mixing flows. The use of turbulence models required at higher Reynolds numbers could reduce the sensitivity of the model to transient buoyancy effects induced by flow instabilities. The hypothesis is that understanding the physics first in the laminar flow regime will facilitate the development of future turbulent models with additional physics and more realistic assumptions.

In the development of the mathematical model, a number of additional simplifying assumptions have been made, namely:
\begin{enumerate}
	\item Density variations are sufficiently small, $\Delta \rho/\rho \ll 1$, for the Boussinesq approximation to be applicable for describing the flow in the tanks. \label{as_incompressible}
    \item The deformation of the electrolyte free surface caused by the impact of the discharging jet is considered negligible. \label{as_freesurface}
	\item The flow in the pipes and the cell is adiabatic, meaning that no heat is transferred to or from the environment. \label{as_adiabatic}
    \item The inner walls of the tanks are kept at ambient temperature, $T_{\textrm{amb}}$. \label{as_walls}
	\item The heat transfer between the anolyte and catholyte is perfect as they flow through the cell, ensuring that both leave the cell at the same temperature. \label{as_uniformT}
	\item The electrolytes become homogeneously mixed when they flow through the half-cells, so they enter the tank with uniform composition. \label{as_uniformSoC}
    \item The crossover of vanadium ions through the membrane is neglected. \label{as_nocrossover}
\end{enumerate}

Assumption \ref{as_freesurface} is a good approximation as long as the dynamic pressure of the incoming jet, $\rho u_{\textrm{in}}^2/2$, is small compared to the hydrostatic pressure, $\rho g D_{\textrm{P}}$, associated with (vertical) free surface disturbances of the order of $D_{\textrm{P}}$. This implies that the characteristic Froude number remains small or of order unity in all cases, being $\textrm{Fr}=u_\textrm{in}^2/(gD_\textrm{P})\in [3\cdot 10^{-3}, 1.41]$ in all cases considered in the study. Since $D_\textrm{P}\ll L_{\rm tank}$, the deformation of the free surface is thus negligible compared to the characteristic size of the tanks and can be safely neglected in all cases. As a result, the upper boundary of the fluid in the tank can be considered in first approximation a stress-free horizontal free surface that remains at a constant vertical level (Figures \ref{figure_scheme_tank}b and \ref{figure_boundary_conditions}).

Assumption \ref{as_adiabatic} implies that the thermal resistance of the electrochemical cell and the piping system is so high, and the residence times is so low, that no heat is lost to or gained from the environment. By contrast, assumption \ref{as_walls} stems from the large residence times of the electrolytes in the tanks, which for the typically thin walls used in conventional RFB tank designs allows the inner wall temperature to approach that of the surrounding air in first approximation. It is interesting to note that that these two assumptions may partially compensate each other, resulting in a lesser impact on the results when considered together.

Assumptions \ref{as_uniformT} and \ref{as_uniformSoC} allow the use of uniform species concentrations and temperature in the discharging jets, as shown in Figure \ref{figure_boundary_conditions}. Finally, assumption \ref{as_nocrossover} is used to simplify the cell model as much as possible in order to focus the attention on the modeling of the tanks, ignoring cross-contamination effects and the impact of self-discharge reactions.

\subsection{Tank Model}
\label{section_tankmodel}

\subsubsection{Governing equations}

The fluid dynamics of the electrolytes in the tanks is governed by the Navier-Stokes equations
\begin{align}
	\nabla \cdot \vec{u} & = 0  \label{eq_mass} \\
	\rho_0 \frac{\partial \vec{u}}{\partial t} + \rho_0 \left(\vec{u} \cdot \nabla \right) \vec{u} & = -\nabla p + \nabla \cdot \bar{\tau} + \rho \vec{g} \label{eq_momentum}
\end{align}
supplemented by the mass conservation equation for the chemical species
\begin{equation}
	\frac{\partial c_i}{\partial t} + \vec{u} \cdot \nabla c_i = D_i \Delta c_i \quad \text{for} \quad i=\{{\sf II},{\sf III},{\sf IV},{\sf V}\}, \label{eq_dif}
\end{equation}
and the energy equation
\begin{equation}
	\frac{\partial T}{\partial t} + \vec{u} \cdot \nabla T = \frac{k}{\rho_0 c_p} \Delta T \label{eq_energy_T}
\end{equation}
In the above equations, $\vec{u}$ is the fluid velocity, $p$ is pressure, $c_i$ is the molar concentration of species $i$, $T$ is temperature, and $\vec{g} = -g\vec{e}_y$ is the acceleration of gravity. The state of charge of electrolyte $j = \{+,-\}$, hereafter denoted as $\textrm{SoC}^j$, is defined as the ratio of the molar concentrations of charged vanadium ions to the total concentration of vanadium ions in each electrolyte, namely
\begin{equation}
	\textrm{SoC}^- = \frac{c_{\sf II}}{c_{\sf II}+c_{\sf III}} \quad \textrm{and} \quad \textrm{SoC}^+ = \frac{c_{\sf V}}{c_{\sf IV}+c_{\sf V}}, \label{eq_SoC}
\end{equation}
where, since crossover effects are neglected (assumption \ref{as_nocrossover}), the total vanadium concentration
\begin{equation}
c_\textrm{tot} = c_{\sf II}+c_{\sf III} = c_{\sf IV}+c_{\sf V}
\label{eq_c_tot}
\end{equation}
is the same for both tanks and remains constant over time. Following standard practice, the four oxidation states of vanadium V$^{2+}$, V$^{3+}$, VO$^{2+}$ and VO$_2^+$ are denoted here as V$^{{\sf II}}$, V$^{{\sf III}}$, V$^{{\sf IV}}$, and V$^{{\sf V}}$, or, in abbreviated form, with subindex $i=\{{\sf II},{\sf III},{\sf IV},{\sf V}\}$ as in equation~\eqref{eq_dif}.

In the above equations, the fluid density is assumed to be constant and equal to the reference density, $\rho_0$, except in the buoyancy term of equation \eqref{eq_momentum}, where $\rho$ is assumed to be a linear function of $T$ and SoC (see section \ref{sec_density}). Moreover, $\bar{\tau} = \mu [ \nabla \vec{u} + {\left( \nabla \vec{u} \right)}^T ]$ is the viscous stress tensor, written in terms of the fluid viscosity $\mu$, assumed also to be dependent on $T$ and SoC (see section \ref{sec_viscosity}). The equations presented above are strongly coupled, as the motion of the electrolyte drives the convective transport of temperature and active species, while the resulting temperature and concentration fields modify the electrolyte density and viscosity, thus affecting the flow and closing the loop.

As implied by assumption \ref{as_incompressible}, when writing equation \eqref{eq_momentum} the density changes are considered so small that they are only important in the buoyancy force. The Boussinesq approximation relies on the use of a variable density function in the gravitational term, $\rho \vec{g}$, but a constant (i.e., reference) density, $\rho_0$, in all other terms of the conservation equations. In thermal buoyancy problems these small density variations are expressed as $\rho=\rho_0\left[1-\beta (T-T_0) \right]$ in terms of the thermal expansion coefficient $\beta=-\rho^{-1}\left( \partial \rho / \partial T \right)_p$ at the reference temperature $T_0$. As mentioned in section \ref{section_intro}, in our problem the density of the electrolytes changes both with $T$ and $\textrm{SoC}$, so the density function must be generalized as done in section \eqref{sec_density} below to account for both variables simultaneously.

The transport of species $i=\{{\sf II},{\sf III},{\sf IV},{\sf V}\}$ is described by the convection-diffusion equation \eqref{eq_dif}. The diffusion coefficients $D_i$ are assumed to be independent of temperature and viscosity, with their values adopted from \cite{Munoz2022}. It is interesting to note that, due to the low diffusivities of the active vanadium species, the Peclet number for mass diffusion takes extremely large values in all cases under study, with values in the range $\textrm{Pe}=D_{\textrm{P}}u_\textrm{in}/D_i \in [2\cdot 10^6, 5\cdot 10^7]$. This means that convective transport dominates over diffusive transport, which leads to the formation of thin mixing layers with strong concentration gradients that impose strict numerical constraints on the size of the computational cells.

As implied by assumption \ref{as_nocrossover}, the crossover of vanadium species through the membrane and the associated cross-contamination effects and self-discharge reactions \cite{Tang2012,Ibanez2021} are considered to have a small effect in the concentration and energy balances \eqref{eq_dif} and \eqref{eq_energy_T}, so they are neglected in first approximation. The heat released by the self-discharge reactions is also neglected for consistency. The energy equation \eqref{eq_energy_T} does not include the viscous dissipation term either, which is also expected to be small. Regarding the thermal conductivity, $k$, and specific heat, $c_p$, correlations of the latter with $T$ for $\rm VOSO_4\cdot2.63\ H_2O(s)$ were tabulated by Quin et~al.~\cite{Qin2010}, who showed a weak dependence at small temperature ranges. To the authors' knowledge, there is no available data on the dependency of $k$ and $c_p$ with $\textrm{SoC}$, so both parameters will be assumed to be constant due to the small temperature variations expected in the tanks, an assumption that is commonly accepted in the literature \cite{Tang2012-2,Tang2012,Xiong2013,Wei2014,We2014,Yan2016}.

\subsubsection{Density as a function of $T$ and {\rm SoC}}

\label{sec_density}

In this study, where small temperature and composition changes are considered, the variation of fluid density with $T$ and SoC is  approximated as linear for both variables
\begin{equation}
	\rho^j = \rho_0^j(T_0) + \rho_T^j (T^j-T_0) + \rho^j_{\textrm{SoC}} \textrm{SoC}^j  \quad j=\{+,-\} \label{eq_rho}
\end{equation}
where $T_0$ is the reference temperature at which the reference density $\rho_0^j$ is known, and the factors $\rho_T^j = (\partial \rho^j/\partial T^j)_{\textrm{SoC}^j}$ and $\rho_{\textrm{SoC}}^j = (\partial \rho^j/\partial \textrm{SoC}^j)_{T^j}$ are the partial derivatives of density with respect to $T$ and SoC for tank $j$. The former are assumed equal for both electrolytes, $\rho^+_T = \rho^-_T = -0.6$~kg~m$^{-3}$~K$^{-1}$ \cite{Mousa2003,Ressel2018}, while the latter are taken as $\rho^+_{\textrm{SoC}}=10$~kg~m$^{-3}$ \cite{Skyllas2016} and $\rho^-_{\textrm{SoC}}=-30$~kg~m$^{-3}$ \cite{Skyllas2016,Ressel2018}, close to the values reported by other authors. The reference densities considered in this work are $\rho_0^+ = 1390$ kg m$^{-3}$ and $\rho_0^- = 1410$ kg m$^{-3}$ \cite{Ressel2018}. For a more comprehensive discussion of the literature data on the density of vanadium electrolytes, the reader is referred to \ref{section_app_data}.

\subsubsection{Viscosity as a function of $T$ and {\rm SoC}}

\label{sec_viscosity}

Li et~al.~\cite{Li2018} reported tabulated data correlating viscosity with $T$ and SoC, while Xu et~al.~\cite{Xu2014} reported correlations between viscosity, SoC, and vanadium concentration. Both studies fitted the experimental data for the positive and negative electrolytes to a second-order polynomial
\begin{equation}
	\mu^j = A^j + B^j \, \textrm{SoC}^j +C^j \, T^j + D^j \, \textrm{SoC}^{j,2} + E^j \, T^j \textrm{SoC}^j + F^j \, T^{j,2} \quad j=\{+,-\} \label{eq_viscosity}
\end{equation}
where coefficients $A^j$ to $F^j$ depend on the total vanadium and sulphate concentrations \cite{Munoz2022}. This work uses the coefficients of Li et~al.~\cite{Li2018} listed in Table \ref{table_viscosity}, which are assumed to remain constant with time. If species and water crossover were considered, this assumption might fail when the two electrolytes become sufficiently unbalanced, as the total vanadium concentration would differ between the tanks. It is also assumed that the difference in sulfate concentration does not affect significantly the value of viscosity, which is in addition a parameter of secondary importance for the flow in the tanks due to the large Reynolds numbers of the flow.

\begin{table}[ht!]
	\centering
	\caption{Coefficients used in the viscosity equation \eqref{eq_viscosity} \cite{Li2018}.\label{table_viscosity}}
	\begin{tabular}{lrr}
		\hline
		Parameter & $+$ Electrolyte & $-$ Electrolyte \rule{0pt}{2.6ex}\\
		\hline
		A (Pa s)& $3.369 \cdot 10^{-1}$ & $5.100 \cdot 10^{-1}$\\
		B (Pa s)& $-2.221 \cdot 10^{-2}$ & $-4.438 \cdot 10^{-2}$\\
		C (Pa s K$^{-1}$)& $-2.043 \cdot 10^{-3}$ & $-3.077 \cdot 10^{-3}$\\
		D (Pa s)& $5.906 \cdot 10^{-4}$ & $-1.103 \cdot 10^{-4}$\\
		E (Pa s K$^{-1}$)& $6.627 \cdot 10^{-5}$ & $1.358 \cdot 10^{-4}$\\
		F (Pa s K$^{-2}$)& $3.127 \cdot 10^{-6}$ & $4.679 \cdot 10^{-6}$\\
		\hline
	\end{tabular}
\end{table}

\begin{figure}[b!]
	\centering
	\includegraphics{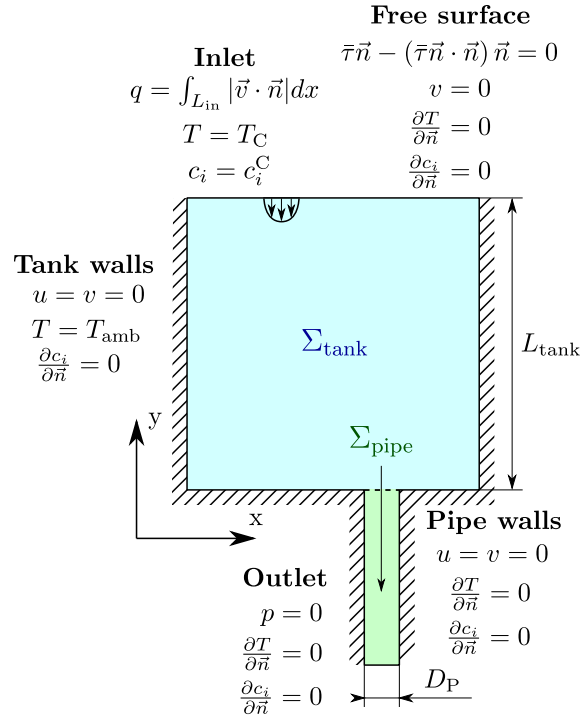}
	\caption{Sketch of the tank indicating the boundary conditions applying over each domain boundary.\label{figure_boundary_conditions}}
\end{figure}

\subsubsection{Boundary conditions}

Figure \ref{figure_boundary_conditions} summarizes the boundary conditions used for the numerical integration of the problem. The free surface between the electrolyte and the gas in the tank is assumed to be adiabatic and impermeable to vanadium species. In addition, we impose a zero vertical velocity condition and a vanishing normal projection of the stress tensor to ensure that no vertical velocities appear in this boundary, as implied by assumption \ref{as_freesurface}. Uniform temperature and species concentrations are imposed at the inlet section (assumptions \ref{as_uniformT} and \ref{as_uniformSoC}) where a fully developed laminar flow is imposed, with the same 2D volume flow rate per unit length
\begin{equation}
q = \int_{L_{\textrm{in}}} |\vec{v} \cdot \vec{n}| \, {\textrm{d}}x
\end{equation}
for both tanks, where $L_{\textrm{in}}$ refers to the inlet boundary which has a width $D_{\rm P}$ in the $x$-direction, just like the outlet pipe. The value of $q$ is computed using the 3D flow rate, $Q$, used in the cell model presented below so as to preserve the Reynolds number in the 2D tank simulations and thus obtain qualitatively relevant results. For the conversion, the Reynolds number $\textrm{Re}_\textrm{2D} = \rho q/\mu$ based on the 2D flow rate $q$ is assumed to be equal to the Reynolds number $\textrm{Re}_\textrm{3D} = \rho (Q/A) D_{\rm P}/\mu$ of a 3D jet with volumetric flow rate $Q$ that flows from a pipe of diameter $D_\textrm{P}$ and cross-sectional area $A = \pi D_{\rm P}^2/4$. This condition yields the relation
\begin{equation}
     q = \frac{4 Q}{\pi D_\textrm{P}}
\end{equation}
that provides $q$ in terms of $Q$. The inlet values for $T$ and $c_i$ are obtained from the outlet values of the cell model. A constant temperature $T_{\textrm{amb}}$ is imposed at the tank walls (assumption \ref{as_walls}) as well as the no-slip boundary condition and the zero flux boundary conditions for all species. Finally, a constant reference pressure is imposed at the outlet of the discharging pipe, that is placed sufficiently far downstream the tank outlet section to avoid that any upstream perturbation may affect the tank flow.

The species concentrations and temperature at the outlet of the tanks, which are affected by the mixed convection flow of the electrolytes in the tanks, influence the electrochemical cell since they are the inputs to the cell model via the outlet sections. These values are computed from their respective convective flows, with the averaged outlet concentration of species $i$ given by
\begin{equation}
    \overline{c^{\textrm{out}}_i} = \frac{1}{q} \int_{L_{\textrm{out}}} c_i \left( \vec{v} \cdot \vec{n}\right) \, {\textrm{d}}x \quad \textrm{for} \quad i=\{{\sf II},{\sf III},{\sf IV},{\sf V}\}, \label{eq_cout}
\end{equation}
and the averaged outlet temperature of electrolyte $j$ by
\begin{equation}
    \overline{\rho^j T_{\textrm{out}}^j} = \frac{1}{q} \int_{L_{\textrm{out}}} \rho^j T^j\left( \vec{v} \cdot \vec{n}\right) \, {\textrm{d}}x \quad j=\{+,-\}, \label{eq_rhoTout}
\end{equation}
where $\vec{n}$ denotes the outward unit normal and $L_{\rm out}$ refers to the outlet boundary, whose width $D_{\rm P}$ in the $x$-direction is assumed to be equal to that of the inlet sections.

\subsection{Cell Model}\label{section_cellmodel}


The model used for the redox flow battery cell is zero-dimensional and is linked to the tank model by assuming that the residence time in the pipes and the cell is much shorter than the residence time in the tanks. This means that any processes occurring outside of the tanks are considered to happen instantaneously. As a result, the fluid coming out of the tanks is immediately reintroduced through the inlet sections  slightly different properties due to the electrochemical reactions taking place in the cell. The variables of the cell, such as species concentrations and electrolyte temperatures, are evaluated at the cell outlet and are denoted with a superscript C (if there is a single value common for both half-cells, e.g., temperature, or there is no possible confusion, e.g., species concentrations) or HC (if both half cells may exhibit different values, e.g., density or SoC). These variables are assumed to be uniform throughout the cell and equal at the inlets of both tanks (assumptions \ref{as_uniformT} and \ref{as_uniformSoC}).

\subsubsection{Chemical reactions in the cell}
The variation of the species concentrations between the tank outlet and inlet are due to the electrochemical cell reactions
\begin{align}
	\textrm{VO}^{2+} + \textrm{H}_2\textrm{O} & \xrightleftharpoons[\textrm{discharge}]{\textrm{charge}} \textrm{VO}_2^+ + 2\textrm{H}^+ + \textrm{e}^- \label{reac_pos} \\
	{\rm V}^{3+} + \textrm{e}^- & \xrightleftharpoons[\textrm{discharge}]{\textrm{charge}} {\rm V}^{2+} \label{reac_neg}
\end{align}
resulting in the following active species concentrations at the cell outlet
\begin{subequations}
\label{eq_ciC}
\begin{align}
	 c_{\textrm{i}}^{\textrm{C}} &= \overline{c^{\textrm{out}}_i} + \frac{I}{FQ} \quad \textrm{for} \quad i=\{{\sf II},{\sf V}\}, \label{eq_cVcharged} \\
	 c_{\textrm{i}}^{\textrm{C}} &= \overline{c^{\textrm{out}}_i} - \frac{I}{FQ} \quad \textrm{for} \quad i=\{{\sf III},{\sf IV}\}, \label{eq_cVdischarged}
\end{align}
\end{subequations}
for the charged and discharged species, respectively. Here $I$ is the total applied current, considered positive during charge and negative during discharge, $F = 96\,485$ C/mol is Faraday's constant, and $Q$ is the volume flow rate, assumed equal for both half cells. The SoC of electrolyte $j$ at the half cell outlet, $\textrm{SoC}_\textrm{HC}^j$, is obtained using equation \eqref{eq_SoC} with the concentration of the corresponding vanadium ions at the cell outlet given in \eqref{eq_ciC}.

\subsubsection{Energy balance in the cell}

The overall energy balance for the cell can be written as follows
\begin{equation}
	\Delta U^+ + \Delta U^- = \dot{Q}_{\textrm{J}} + \dot{Q}_{\textrm{r}}, \label{eq_energy_balance}
\end{equation}
where
\begin{equation}
    \Delta U^j = Q c_p \left(\rho^j_{\textrm{HC}} T_{\textrm{C}} - \overline{\rho^j T_{\textrm{out}}^j} \right)
    \label{eq_deltaU_j} \quad j=\{+,-\}
\end{equation}
is the variation of the internal energy of electrolyte $j$ between the half-cell inlet and outlet, $\dot{Q}_{\textrm{J}}$ is the heat added to the electrolyte by Joule heating per unit time, and $\dot{Q}_{\textrm{r}}$ is the heat evolved by the electrochemical reactions per unit time. Solving the above equations for the cell temperature gives
\begin{equation}
    T_{\textrm{C}} = \frac{1}{\rho^+_{\textrm{HC}} + \rho^-_{\textrm{HC}}} \left( \overline{\rho^+ T_{\textrm{out}}^+} + \overline{\rho^- T_{\textrm{out}}^-} + \frac{\dot{Q}_{\textrm{J}} + \dot{Q}_{\textrm{r}}}{Q c_p} \right)\label{eq_TC}
\end{equation}
which retains the density variations due to the changes in $T$ and $\textrm{SoC}$ that occur the cell, as they are relevant for the energy balance and therefore for the determination of the inlet temperature $T_{\textrm{C}}$. The heats $\dot{Q}_{\rm J}$ and $\dot{Q}_{\rm R}$ transferred to the electrolytes in the cell can be expressed as
\begin{align}
    \dot{Q}_{\textrm{J}} &= I^2 R_{\textrm{C}} \label{eq_Qj}\\
    \dot{Q}_{\textrm{R}} &= \frac{I}{F} \left(\overline{T_{\textrm{out}}^+} \Delta S^+ + \overline{T_{\textrm{out}}^-} \Delta S^-\right) \label{eq_Qr}
\end{align}
in terms of the cell ohmic resistance $R_{\textrm{C}}$ and the molar reaction entropies at standard conditions $\Delta S^+ = \Delta S_{\textrm{VO}^{2+}} + \Delta S_{\textrm{H}_2\textrm{O}} - \Delta S_{\textrm{VO}_2^+} - 2\Delta S_{\textrm{H}^+}$ and $\Delta S^- = \Delta S_{{\rm V}^{3+}} - \Delta S_{{\rm V}^{2+}}$ listed in Table \ref{table_data_Vanadio}. Notice that the evaluation of the entropic heats in equation \eqref{eq_Qr} involves the use of the approximate averaged outlet temperatures $\overline{T_{\textrm{out}}^j}=\overline{\rho^j T_{\textrm{out}}^j}/\rho^j_{\textrm{HC}}$. Other proxies for the average cell temperature, such as $T_{\textrm{C}}$ or the arithmetic mean between the inlet and outlet temperatures, would result in an implicit non-linear equation for $T_{\textrm{C}}$ that would require an iterative solution.

It is important to note that, to retain temperature variations in the tanks, this work employs a highly simplified cell model. In addition to ignoring the heat exchanged with the environment (assumption \ref{as_adiabatic}) and the pump heating, equation \eqref{eq_energy_balance} also ignores the source terms associated with activation losses, HSO$_4^-$ dissociation, and self-discharge reactions \cite{Munoz2023}. All these terms have the capacity to change both the magnitude of temperature variations and their sign, as they may take very different values depending on the cell potential, both in charge and discharge, and on overall system design. In small scale systems, with temperature changes of barely one or two degrees Celsius, the variations in density due to temperature have a minor effect on the flow field compared to those associated with the SoC  (see section \ref{section_dimensional_variations} for details), so the results are expected to be remain unaffected. However, future extensions of this model to industrial scale VRFBs should retain these terms, as the larger temperature changes in the cell and piping system will surely have a more significant impact on the fluid dynamics of the tanks.

\begin{table}[t!]
	\centering
	\caption{Physical properties of the different species involved.\label{table_data_Vanadio}}
	\begin{tabular}{crc}
		\hline
		Parameter & Value & Reference \rule{0pt}{2.6ex}\\
		\hline
		$\Delta S_{{\rm V}^{2+}}$ & $-130$ (J mol$^{-1}$ K$^{-1}$)& \cite{Wei2014} \\
		$\Delta S_{{\rm V}^{3+}}$ & $-230$ (J mol$^{-1}$ K$^{-1}$)& \cite{Wei2014} \\
		$\Delta S_{\textrm{VO}^{2+}}$ & $-133.9$ (J mol$^{-1}$ K$^{-1}$)& \cite{Wei2014} \\
		$\Delta S_{\textrm{VO}_2^+}$ & $-42.3$ (J mol$^{-1}$ K$^{-1}$)& \cite{Wei2014} \\
		$\Delta S_{\textrm{H}^+}$ & $0$ (J mol$^{-1}$ K$^{-1}$)& \cite{Wei2014} \\
		$\Delta S_{\textrm{H}_2\textrm{O}}$ & $69.9$ (J mol$^{-1}$ K$^{-1}$)& \cite{Wei2014} \\
		$\Delta S^+$ & $-21.7$ (J mol$^{-1}$ K$^{-1}$)& - \\
		$\Delta S^-$ & $-100$ (J mol$^{-1}$ K$^{-1}$)& - \\
		$D_{{\sf II}}$ & $8.10\cdot 10^{-12}$ (m$^2$ s$^{-1}$)& \cite{Munoz2022} \\
		$D_{{\sf III}}$ & $1.65\cdot 10^{-11}$ (m$^2$ s$^{-1}$)& \cite{Munoz2022} \\
		$D_{{\sf IV}}$ & $2.37\cdot 10^{-11}$ (m$^2$ s$^{-1}$)& \cite{Munoz2022} \\
		$D_{{\sf V}}$ & $4.16\cdot 10^{-12}$ (m$^2$ s$^{-1}$)& \cite{Munoz2022} \\
		\hline
	\end{tabular}
\end{table}

\subsubsection{Cell potential}

The equilibrium potentials for the positive and negative electrodes are written as follows
\begin{subequations}
\begin{align}
    E_\textrm{eq}^+ &= E_0^+ + \frac{RT_\textrm{C}}{F} \ln{\left(\frac{c_{\sf V}^\textrm{C}}{c_{\sf IV}^\textrm{C}}\right)} \label{eq_E_+}\\
    E_\textrm{eq}^- &= E_0^- + \frac{RT_\textrm{C}}{F} \ln{\left(\frac{c_{\sf III}^\textrm{C}}{c_{\sf II}^\textrm{C}}\right)} \label{eq_E_-}
\end{align}
\end{subequations}
where $E_0^+$ and $E_0^-$ are the corresponding formal potentials and $R$ is the universal gas constant. When combined with a linear voltage drop associated with ohmic losses, the above expressions provide the following polarization relation between the cell potential $E$ and the applied current $I$
\begin{equation}
    E = E_0 + \frac{RT_\textrm{C}}{F} \ln{\left(\frac{c_{\sf II}^\textrm{C}c_{\sf V}^\textrm{C}}{c_{\sf III}^\textrm{C}c_{\sf IV}^\textrm{C}}\right)} + IR_\textrm{C} \label{eq_E_curve}
\end{equation}
a simplified expression that ignores activation and mass transfer losses and membrane potential drop, but is used here as first approximation to compute the cell potential and thus allow to stop the battery operation when a preset cut-cell potential is reached. In the above expression $E_0=E_0^+-E_0^-$ is the formal cell potential and $R_\textrm{C}$ represents the overall cell ohmmic resistance.

\section{Order of magnitude estimates} \label{section_dimensional_analysis}

This section presents an order-of-magnitude analysis with the aim to anticipate the nature of the mixed-convection flow in the tanks of VRFBs. These estimates are based on the electrolyte properties and operational variables, and are based on the Richardson number Ri defined in \eqref{eq_Ri_intro}. The discussion begins by estimating the variations of the electrolyte properties in a single cell pass, and continues with a detailed discussion of the transition between the flow regimes dominated by momentum, $|\rm Ri| \ll 1$, and buoyancy, $|\rm Ri| \gg 1$.

\subsection{Variation of the electrolyte properties in a single cell pass}

\label{section_dimensional_variations}

As the electrolytes flow through the cell, the electrochemical reactions cause changes in their composition, which in turn affect their physical properties. The most significant change affecting the flow in the tanks concerns density, whereas other properties such as viscosity or specific heat either have a minor role or remain virtually unchanged. One can estimate the relative variations in density by rewriting equation \eqref{eq_rho} for the density as a function of $T$ and SoC in the form
\begin{equation}
	\frac{\Delta \rho^j}{\rho_0^j} = \frac{\rho_T^j\Delta T^j}{\rho_0^j} + \frac{\rho_{\textrm{SoC}}^j \Delta\textrm{SoC}^j}{\rho_0^j} \quad j = \{+,-\} \label{eq_Deltarho}
\end{equation}
where $\Delta T^j$ and $\Delta\textrm{SoC}^j$ denote the single pass variations of temperature and state of charge of electrolyte $j$ as it flows through the cell. It is worth noting that when crossover effects are ignored both electrolytes suffer identical variations of SoC, since from \eqref{eq_SoC}, \eqref{eq_c_tot}, and \eqref{eq_cVcharged} we have
\begin{equation}
    \Delta {\rm SoC}^- = \frac{\Delta c_{\rm II}}{c_{\rm tot}} = \frac{\Delta c_{\rm V}}{c_{\rm tot}} = \Delta {\rm SoC}^+
\end{equation}
Thus, hereafter we should use $\Delta {\rm SoC}$ to denote $\Delta {\rm SoC}^-$ and $\Delta {\rm SoC}^+$ indistinctly. Finally, from equations \eqref{eq_cVcharged} and \eqref{eq_cVdischarged} and the definition of the total vanadium concentration given in \eqref{eq_c_tot}, $\Delta\textrm{SoC}$ can be written as
\begin{equation}
	\Delta \textrm{SoC} = \frac{I}{FQc_\textrm{tot}} \label{eq_DeltaSoC}
\end{equation}

The variation of the electrolyte temperature in a single pass, $\Delta T \sim \Delta T^- \sim \Delta T^+ \ll T_0$, can be estimated from the overall energy balance \eqref{eq_energy_balance}. To this end, it is convenient to introduce the following approximated notation in \eqref{eq_Qr}
\begin{equation}
    \overline{T_{\textrm{out}}^+} \Delta S^+ + \overline{T_{\textrm{out}}^-} \Delta S^- \simeq ~T_0 (\Delta S^+ + \Delta S^-)
\end{equation}
and to estimate the change in the internal energy of the electrolyte in \eqref{eq_energy_balance}-\eqref{eq_deltaU_j} as follows
\begin{equation}
    \Delta U^j \simeq Q c_p \Delta {\left( \rho T \right)}^j = Q c_p T^j \rho^j \left( \frac{\Delta T^j}{T^j} + \frac{\Delta \rho^j} {\rho^j} \right) \simeq Q c_p T_0 \frac{\left(\rho_0^+ + \rho_0^-\right)}{2} \left( \frac{\Delta T}{T_0} + \frac{\Delta \rho^j} {\rho_0^j} \right).
\end{equation}
where we have written $T^j\simeq T_0$, and approximated $\rho^j\simeq \rho^j_0 \simeq (\rho^+_0 + \rho^-_0)/2$ with small errors, according to assumption \ref{as_incompressible}. Using \eqref{eq_Deltarho} and \eqref{eq_DeltaSoC}, the above expression becomes
\begin{equation}
    \Delta U^j \simeq Q c_p T_0 \frac{\left(\rho_0^+ + \rho_0^-\right)}{2} \left( \frac{\Delta T}{T_0} + \frac{\rho_T^j \Delta T}{\rho_0^j} + \frac{\rho_{\textrm{SoC}}^j }{\rho_0^j} \frac{I}{FQc_\textrm{tot}}\right)
\end{equation}
which, upon substitution in the cell energy balance  \eqref{eq_energy_balance}, leads to
\begin{multline}
    Q c_p \left(\rho_0^+ + \rho_0^-\right) \left\{\left[1 + \dfrac{T_0}{2} \left(\dfrac{\rho_T^+}{\rho_0^+}+\dfrac{\rho_T^-}{\rho_0^-}\right)\right]\Delta T + \dfrac{T_0}{2}\left( \dfrac{\rho_\textrm{SoC}^+}{\rho_0^+} + \dfrac{\rho_\textrm{SoC}^-}{\rho_0^-} \right)\dfrac{I}{FQc_\textrm{tot}} \right\} \\ \simeq I^2 R_{\textrm{C}} + \frac{I}{F} T_0 (\Delta S^+ + \Delta S^-)
\end{multline}
finally providing
\begin{equation}
	\Delta T \simeq \frac{\dfrac{I^2R_C + IT_0(\Delta S^+ + \Delta S^-)/F }{Qc_p\left(\rho^+_0 + \rho^-_0 \right)} - \dfrac{T_0}{2}\left( \dfrac{\rho_\textrm{SoC}^+}{\rho_0^+} + \dfrac{\rho_\textrm{SoC}^-}{\rho_0^-} \right)\dfrac{I}{FQc_\textrm{tot}}}{1 + \dfrac{T_0}{2} \left(\dfrac{\rho_T^+}{\rho_0^+}+\dfrac{\rho_T^-}{\rho_0^-}\right)} \label{eq_DeltaT}
\end{equation}

In the above expression, the first term in the numerator represents the combined contribution of Joule heating and the entropic heat of reaction, whereas the second term represents the effect of density changes induced by $\textrm{SoC}$. These two terms compete with each other, and depending on whether the cell is in charge or discharge, and the sign and absolute values of $\rho_\textrm{SoC}^j$, the temperature variation $\Delta T$ may be positive or negative under different operating conditions. It should be noted that the second term in the denominator, representing the effect of density variations due to changes in temperature, has a small effect (ca.~12\%) in terms of the absolute temperature variations observed in the cell, and therefore could be neglected in first approximation. However, we have retained it here for completeness.

Figure \ref{figure_Ri} shows $\Delta T$ isocontours in the $(I, Q)$-plane as obtained from \eqref{eq_DeltaT} using the electrolyte parameters given in tables \ref{table_data_Vanadio} and \ref{table_parameters}. As can be seen, for sufficiently large charge and discharge currents cell heating may grow from tenths to tens of Celsius as the flow rate is decreased, whereas cell cooling is limited to a few degrees Celsius at low to intermediate charge currents and very small flow rates, produced by the entropic heat absorption. Apart from the trivial solution, there is a particular charge current that results in zero temperature change, namely $I_{\Delta T=0}$ with the two heat terms $Q_\textrm{R}$ and $Q_\textrm{J}$ competing and cancelling each other. However, the cell cooling and $I_{\Delta T=0}$ may be produced at different currents or even be neglected if more heat sources are considered, as detailed in section \ref{section_cellmodel}.


\subsection{Buoyancy \textit{vs.} momentum: The Richardson number}
\label{section_dimensional_Richardson}

\begin{figure}[t!]
	\centering
	\includegraphics{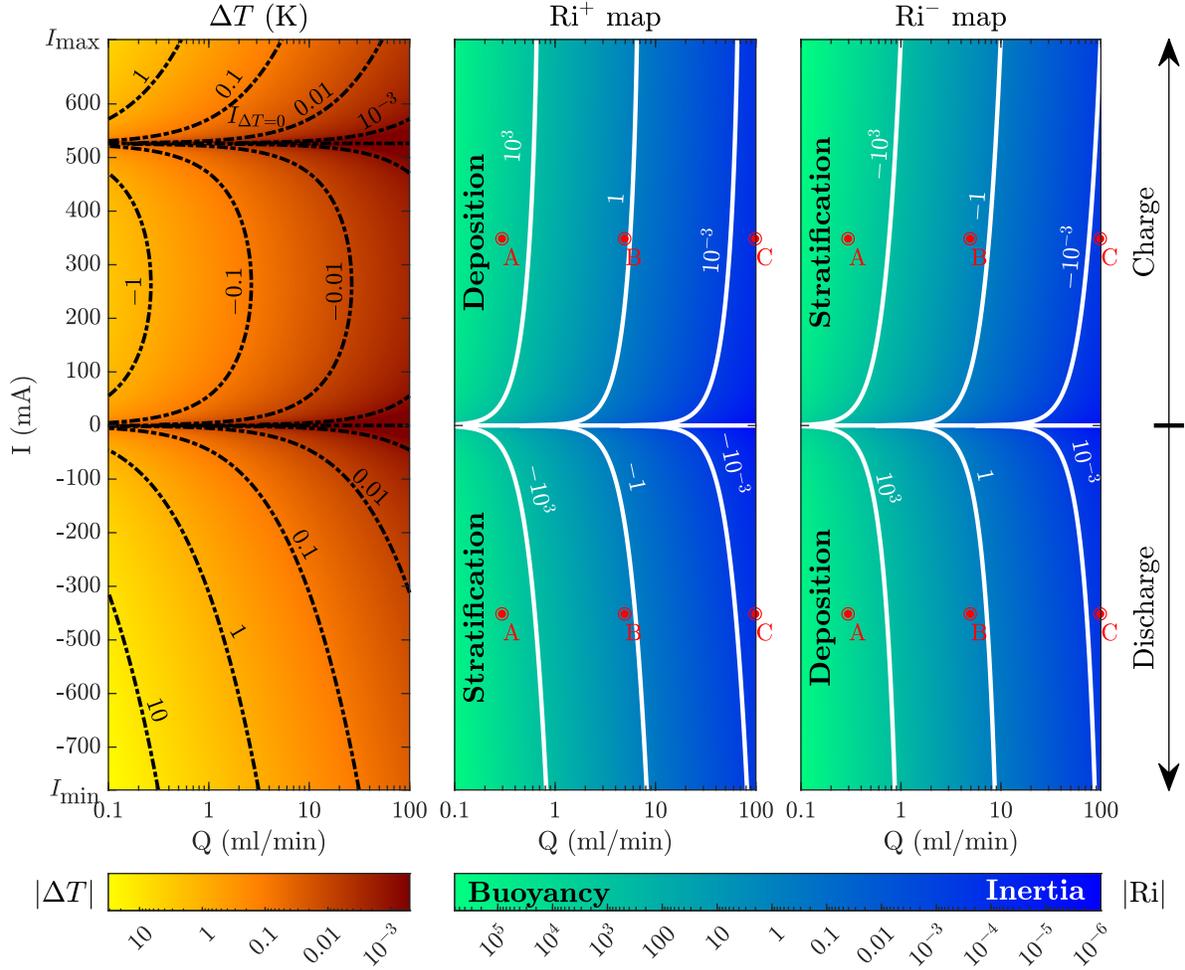}
	\caption{Isocontour maps of $\Delta T$, $\textrm{Ri}^+$ and $\textrm{Ri}^-$ in the $(I,Q)$-plane as obtained from equations \eqref{eq_DeltaT} and \eqref{eq_Ri} with use made of the electrolyte parameters listed in tables \ref{table_data_Vanadio} and \ref{table_parameters}. The regimes of strong stratification (negatively buoyant jets) and fast deposition of the renewed electrolyte (positively buoyant jets) that emerge in buoyancy dominated flows $|\textrm{Ri}^j| \gg 1$ are indicated for reference purposes. \label{figure_Ri}}
\end{figure}

The mixing flow of the electrolytes in the tanks is determined by the competition between the buoyancy forces due to density variations and the momentum flux of the incoming jet. As previously discussed, this competition is characterized by the value of the Richardson number. Once the variations of SoC and $T$ have been obtained from equations \eqref{eq_DeltaSoC} and \eqref{eq_DeltaT}, density variations can be evaluated from equation \eqref{eq_Deltarho}, resulting in the following approximate expression for the Richardson number
\begin{multline}
	\textrm{Ri}^j = \frac{gL_\textrm{tank}}{u^2_\textrm{in}} \frac{\Delta \rho^j}{\rho_0^j} = \frac{\pi^2}{16}\frac{gL_\textrm{tank}D_{\textrm{P}}^4}{Q^2} \left(\frac{\rho_T^j \Delta T^j}{\rho^j_0} + \frac{\rho_{\textrm{SoC}}^j\Delta \textrm{SoC}}{\rho^j_0} \right) \\ \simeq \frac{\pi^2}{16}\frac{gL_\textrm{tank}D_{\textrm{P}}^4}{Q^2} \left(\frac{\rho_T^j}{\rho^j_0} \Delta T + \frac{\rho_{\textrm{SoC}}^j}{\rho^j_0} \frac{I}{FQc_\textrm{tot}}\right) \quad j = \{+,-\}
	\label{eq_Ri}
\end{multline}
where $\Delta T$ is given explicitly by \eqref{eq_DeltaT}.
The above equation shows the high sensitivity of $\textrm{Ri}^j$ to the jet diameter, $D_{\rm P}$, and to the volumetric flow rate, $Q$, but also highlights the importance of having reliable values of the parameters $\rho_{\rm SoC}^j$ and $\rho_{T}^j$ in order to anticipate the type of flow established in the tanks. Figure \ref{figure_Ri} shows isocontours of $\textrm{Ri}^-$ and $\textrm{Ri}^+$ in the $(I,Q)$-plane as obtained from equations \eqref{eq_DeltaT} and \eqref{eq_Ri} for the electrolyte parameters listed in tables \ref{table_data_Vanadio} and \ref{table_parameters}. The lower and upper limits of the total applied current, $I \in [I_{\rm min}, I_{\rm max}]$, correspond to the approximated cut-off potentials in discharge and charge, established, respectively, at $E_\textrm{min}= 1$ V and $E_\textrm{max}=1.7$ V at $\textrm{SoC}=0.5$. The plots also include three operational conditions---A, B and C, both in charge and discharge---that will be further discussed in section \ref{section_results} below.

Different flow regimes can be anticipated depending on the sign and absolute value of $\textrm{Ri}^j$. This can be either negative or positive depending on the sign of $\rho_T^j \Delta T$ and $\rho_{\textrm{SoC}}^j I$, resulting in negatively and positively buoyant jets. For $\textrm{Ri}^j < 0$ (negatively buoyant jets) the electrolyte discharged into the tank tends to ``float'' over the denser bulk fluid. In this case, buoyancy forces reduce the initial momentum of the jet, which may even turn into a plume. By way of contrast, for $\textrm{Ri}^j > 0$ (positively buoyant jets) the jet momentum is favoured by gravity and the jet becomes more stable and tends to flow right to the bottom. These behaviors are modulated by the absolute value of the Richardson number. On one hand, for $|\textrm{Ri}^j| \gg 1$ buoyancy completely dominates the flow, leading to either strong stratification (negatively buoyant jets) or fast deposition of the renewed electrolyte (positively buoyant jets). On the other hand, for $|\textrm{Ri}^j| \ll 1$ the flow is momentum controlled, and the jet dynamics is determined by forced convection, remaining quasi-steady during long discharge periods for the small Reynolds numbers considered here. For $|\textrm{Ri}^j| \sim 1$, between the limits of buoyancy and momentum dominated flows, there exists an intermediate region where both effects are comparable and different flow features may appear in the solution depending on the sign and magnitude of $\textrm{Ri}^j$. As occurs with temperature variations, there exists a non-trivial applied current that results in zero density change and thus makes the Richardson number also equal to zero $I_{\textrm{Ri}^j=0}$, but remains outside the studied range of cut-off potentials, being $I_{\textrm{Ri}^+=0}>I_\textrm{max}$ and $I_{\textrm{Ri}^-=0}<I_\textrm{min}$.


The above discussion is based on single cell pass estimations, which neglect the long-term dynamics of the electrolyte flow in the tanks. Over multiple cell passes, cumulative effects can emerge. Even under momentum-dominated conditions, where $|\textrm{Ri}|^j \ll 1$, slight buoyancy effects can ultimately distort and destabilize the flow, altering the dynamics of mixing after a long quasi-steady induction period. These long-term buoyancy effects, which arise from the accumulation of slight density changes over extended operational times, can induce flow destabilization. But their impact, together with the details of the tank geometry, can only be accurately described using extensive numerical simulations.

\section{Numerical method}\label{section_numerical_method}

The solution to the problem stated in section~\ref{section_model} is obtained numerically using the finite element software COMSOL Multiphysics 5.6. It provides the velocity, pressure, temperature and species concentration fields, as well as the corresponding SoC and fluid density and viscosity distributions in both tanks. Another outcome of the solution is the tank-averaged state of charge
\begin{equation}
\textrm{SoC}^j_\textrm{tank} = \frac{1}{\Sigma_{\textrm{tank}}^j} \int_{\Sigma_{\textrm{tank}}^j} \textrm{SoC}^j \, {\textrm{d}}\sigma \quad j = \{+,-\}
\end{equation}
that can be used to describe the overall charge and discharge process, where $\Sigma_{\textrm{tank}}^j$ denotes both the 2D tank surface area and the associated integration domain.

The conservation equations are discretized using linear rectangular elements in the outlet pipe and linear triangular elements in the tank. The mesh is refined near the walls to adequately capture the viscous boundary layer, but remains fine enough within the tanks to solve the transient mixing layers that appear between the renewed electrolyte jet and the fluid already in the tank. A sensitivity analysis was carried out to ensure that a mesh-independent solution was obtained. The final mesh had $182\,842$ elements per tank, with a maximum element size of $\delta x_{\rm max} = 0.225$ mm ($\delta x_{\rm max}/L_{\rm tank} = 4.5\cdot 10^{-3}$), a minimum element size of $\delta x_{\rm min} = 0.0216$ mm ($\delta x_{\rm min}/L_{\rm tank} = 4.32\cdot 10^{-4}$), and a maximum time step of $0.3$ s, with an estimated relative error of $10^{-3}$. The sensitivity study revealed that doubling the number of elements led to a $0.9\%$ change in $\textrm{SoC}_\textrm{HC}$ during the first $10\%$ of the charge. The maximum differences occurred when an initially quasi-steady momentum dominated jet experienced a buoyancy-induced instability, while for flows initially dominated by buoyancy the changes remained below $0.1\%$.

The hardware used for the simulations included an Intel$^\text{\textregistered}$ Xeon$^\text{\textregistered}$ E5-2695 v2 12-Core Processor 2.4 GHz CPU with 80 GB RAM, and took an average of about 26 hours per charge or discharge cycle. The initial condition was obtained by time-marching the constant density Navier-Stokes equations \eqref{eq_mass} and \eqref{eq_momentum} until a steady-state solution was obtained. Thereafter the cell model was switched on and a transient simulation was carried out in galvanostatic mode, starting from a uniform electrolyte composition with homogeneous $\textrm{SoC} = \textrm{SoC}_0 \ll 1$ at $t = 0$. The simulations run until the cell potential reaches either $E_\textrm{max}$ in charge or $E_\textrm{min}$ in discharge. These are the cut-off potentials above which secondary parasitic reactions (i.e., hydrogen and oxygen evolution) may occur, thus reducing the conversion efficiency of the cell.

The time estimated to completely charge/discharge the tanks is given by (see \ref{section_app_CSTR})
\begin{equation}
    t_{\textrm{end}} = \frac{t_{\textrm{tank+pipe}}}{\Delta{\rm SoC}} =  \frac{(\Sigma_{\textrm{tank}}+\Sigma_{\textrm{pipe}})}{q}\frac{F Q c_{\rm tot}}{I} = \frac{\pi}{4} D_{\rm P}\left(\Sigma_\textrm{tank}+\Sigma_\textrm{pipe}\right)\frac{Fc_\textrm{tot}}{I}
\end{equation}
where $t_{\rm tank+pipe} = (\Sigma_{\rm tank}+\Sigma_{\rm pipe})/q$ denotes the characteristic residence time of the electrolyte in the tank and the outlet pipe, defined in terms of the 2D tank and outlet pipe surface areas, $\Sigma_\textrm{tank}$ and $\Sigma_\textrm{pipe}$, and the 2D volumetric flow rate, $q$, and where $\Delta{\rm SoC}$ is given by equation~\eqref{eq_DeltaSoC}. With the operational parameters presented in the next section (see Table \ref{table_parameters}), $t_\textrm{end}$ is equal to $3142.5$ s. This time is also used to compute the theoretical capacity, $C_\textrm{th}=It_\textrm{end} = 1.10\cdot 10^3$~C, when the battery is operated in galvanostatic mode. Following standard practice, the coulombic efficiency and capacity utilization are thus defined as follows
\begin{subequations}
\begin{equation}
 \textrm{Coulombic Efficiency}= \frac{C_\textrm{D}}{C_\textrm{C}}
\end{equation}
and
\begin{equation}
    \textrm{Capacity Utilization}= \frac{C_\textrm{D}}{C_\textrm{th}}
\end{equation}
\end{subequations}
where $C_\textrm{C}=It_\textrm{C}$ is the actual capacity of the battery that charges in a time $t_\textrm{C}$, and $C_\textrm{D}=It_\textrm{D}$ is the actual capacity of the battery that discharges in a time $t_\textrm{D}$.

The numerical results were compared with those obtained with the Continuous Stirred Tank Reactor (CSTR) model, a zero-dimensional model widely adopted in the literature that assumes perfect mixing ignoring buoyancy effects. This model is often used as first approximation for the description of the tanks (e.g., \cite{Tang2012}) and is briefly reviewed in \ref{section_app_CSTR}.

\subsection{Simulation parameters}

The parameters used in the simulations are listed in Table \ref{table_parameters}. They are based on the modeling and experimental work carried out by Mu\~{n}oz-Perales et~al.~\cite{Munoz2022}. The system under study consisted of a small VRFB cell and two identical tanks (see Figure \ref{figure_scheme_tank}a). The electrode surface area is $9$~cm$^2$, which for a typical current density of $40$~mA~cm$^{-2}$ leads to a total applied current of $I=0.35$~A. The ohmic cell resistance is assumed constant, $R_{\textrm{C}}=0.5$ $\Omega$. In both sides a constant volumetric flow rate $Q$ is used, which varies in the range $Q=\{0.3,100\}$ ml min$^{-1}$ to sweep over different flow regimes and battery responses. The electrolyte is stored in 2D square-shaped tanks with side $L_\textrm{tank}=5$ cm, corresponding to the average electrolyte level in the experimental jars of Mu\~{n}oz-Perales et~al.~\cite{Munoz2022}. The length of the outlet pipe was set to $1.25$ cm, long enough to obtain a fully developed velocity profile at the outlet boundary. The electrolyte has a total concentration of vanadium species $c_\textrm{tot} = 1.8$~M in a sulfuric acid solution with 4.64~M of total sulphates, hence the electrolyte properties are based on that composition. As done in previous work, the specific heat is assumed to be constant and equal for both electrolytes, $c_p=3200$~J~kg$^{-1}$~K$^{-1}$ \cite{Tang2012,Tang2012-2,Xiong2013}.

\begin{table}[ht!]
	\centering
	\caption{Parameters used in the numerical simulations. \label{table_parameters}}
	\begin{tabular}{ccc}
        \hline
		Parameter & Value & Reference \rule{0pt}{2.6ex}\\
		\hline
		$I$ & $0.35$ (A) & - \\
		$L_{\textrm{tank}}$ & $0.05$ (m) & - \\
		$D_{\textrm{P}}$ & $3.175 \cdot 10^{-3}$ (m) & - \\
		$R_{\textrm{C}}$ & $0.5$ ($\Omega$) & - \\
		$c_\textrm{tot}$ & $1.8$ (M) & - \\
		$\rho^+_0$ & $1390$ (kg m$^{-3}$) & \cite{Skyllas2016} \\
		$\rho^-_0$ & $1410$ (kg m$^{-3}$) & \cite{Skyllas2016} \\
		$\rho^+_T$ & $-0.6$ (kg m$^{-3}$ K$^{-1}$) & \cite{Skyllas2016} \\
		$\rho^-_T$ & $-0.6$ (kg m$^{-3}$ K$^{-1}$) & \cite{Mousa2003,Ressel2018,Skyllas2016} \\
		$\rho^+_\textrm{SoC}$ & $10$ (kg m$^{-3}$) & \cite{Skyllas2016} \\
		$\rho^-_\textrm{SoC}$ & $-30$ (kg m$^{-3}$) & \cite{Skyllas2016,Ressel2018} \\
		$c_p$ & $3200$ (J kg$^{-1}$ K$^{-1}$) & \cite{Tang2012-2,Tang2012,Xiong2013} \\
		$k$ & $0.67$ (W m$^{-1}$ K$^{-1}$) & \cite{Al-Fetlawi2009} \\
		$T_{\textrm{amb}}$ & $293$ (K) & - \\
		$T_0$ & $293$ (K) & - \\
		$E_0^+$ & $0.98$ (V) & \cite{Munoz2022} \\
		$E_0^-$ & $-0.36$ (V) & \cite{Munoz2022} \\
		$E_0$ & $1.34$ (V) & - \\
		$E_{\textrm{max}}$ & $1.7$ (V) & - \\
		$E_{\textrm{min}}$ & $1$ (V) & - \\
		$\textrm{SoC}_0$ & 0.01 & - \\
		\hline
	\end{tabular}
\end{table}

\subsection{Mixing Index}

To quantify the results in terms of electrolyte mixing, we use Kramer's mixing index \cite{Lacey1954}
\begin{equation}
    \textrm{M}^j = 1-\frac{s^j}{s_0^j}
\end{equation}
defined in terms of the SoC standard deviation obtained numerically
\begin{equation}
    s^j = \left[\frac{1}{\Sigma_{\textrm{tank}}^j} \int_{\Sigma_{\textrm{tank}}^j} {\left( \textrm{SoC}^j - \textrm{SoC}^j_\textrm{tank} \right)}^2 \, {\textrm{d}}\sigma\right]^{1/2}
\end{equation}
and the standard deviation of the fully unmixed solution, $s_0^j=\left[{\textrm{SoC}^j_\textrm{tank}\left(1- \textrm{SoC}^j_\textrm{tank} \right)}\right]^{1/2}$, in which the electrolyte remains completely unmixed with extreme SoC values of either 0 or 1. The values of $\rm M^j$ range from 0 for the fully unmixed solution to 1 for the perfectly mixed (i.e., homogeneous) solution.

\section{Results and discussion}\label{section_results}

This section presents and discusses the numerical results. Firstly, three extreme values of the Richardson number are studied to delimit the types of flows that appear in the tanks. Next, a couple of intermediate cases are simulated to illustrate situations in which transitions between different types of flows may occur due to the cumulative effect of density changes. Finally, the decrease in battery capacity is investigated in detail and a parametric study of the $\rho_{\rm SoC}^+$ coefficient is presented, the value of which is not yet clear in the literature. The reader is refereed to the video abstract, where several results of \textrm{SoC} maps of tanks over time are summarized.

\subsection{Effect of the Richardson number}

\begin{figure}[t!]
	\centering
	\includegraphics{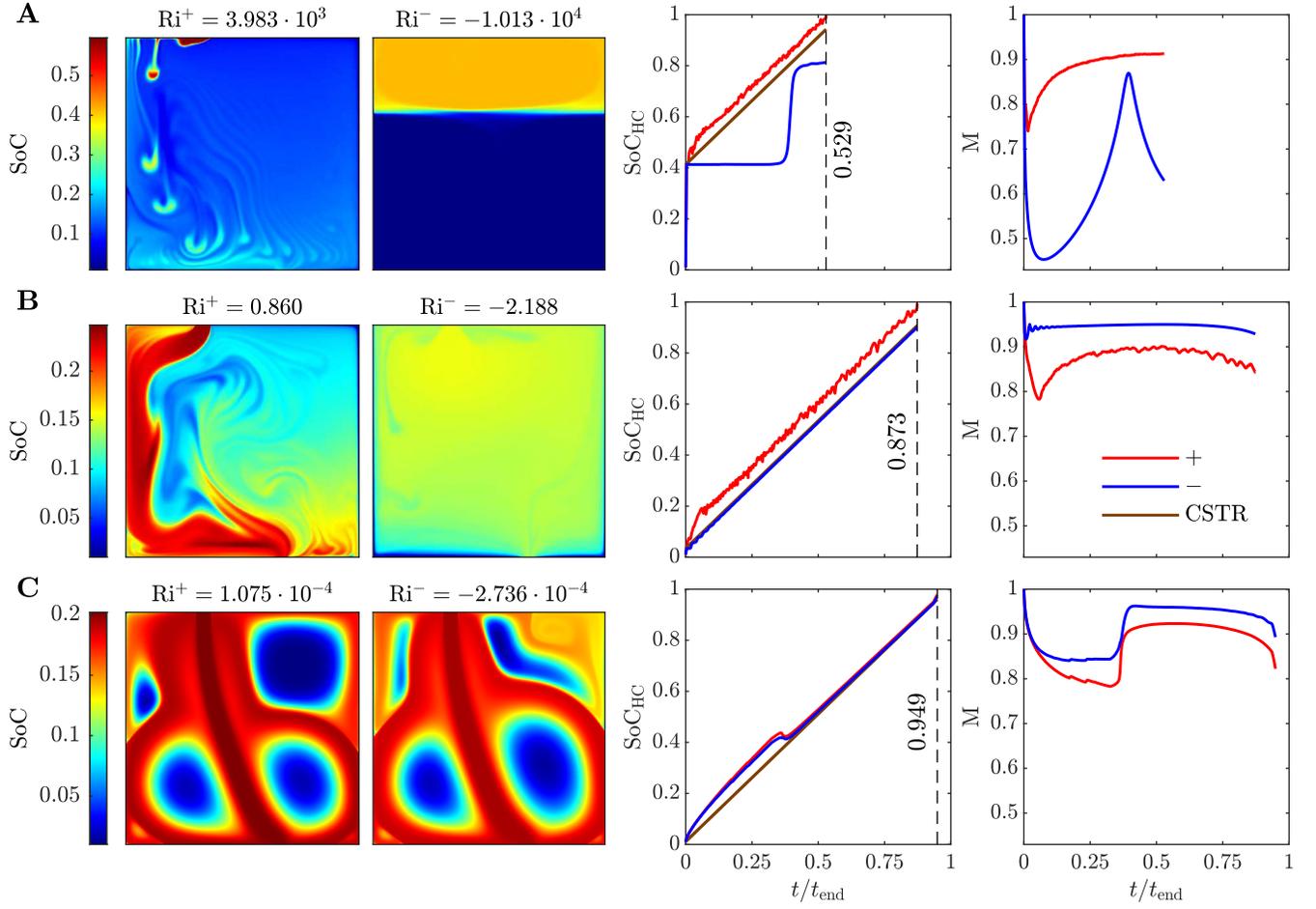}
	\caption{$\textrm{SoC}$ maps for the positive and negative tanks at $t/t_\textrm{end}=0.127$ (left), and time evolution of the average state of charge of both half cells $\textrm{SoC}_\textrm{HC}^j$ and the corresponding mixing index $\rm M^j$ (right) for three volume flow rates $Q = \{0.3,5,100\}$ ml min$^{-1}$ corresponding to charge conditions A, B, and C in Figure \ref{figure_Ri} with $|\textrm{Ri}| \gg 1$, $|\textrm{Ri}| \sim 1$, and $|\textrm{Ri}| \ll 1$, respectively.
 \label{figure_impact_Ri}}
\end{figure}

The aim of this section is to present numerical simulations for the three distinguished limits of the Richardson number, namely $|\textrm{Ri}| \gg 1$, $|\textrm{Ri}| \sim 1$, and $|\textrm{Ri}| \ll 1$. The simulations will be carried out by progressively increasing the volume flow rate in three separate charge cycles. The purpose of this test is to illustrate the different flow regimes that may emerge under the competing effects of momentum and buoyancy. Figure \ref{figure_impact_Ri} shows results obtained for $Q = \{0.3,5,100\}$ ml min$^{-1}$, corresponding to points A, B and C in the upper part of Figure \ref{figure_Ri} ($I>0$, charge). Figure \ref{figure_impact_Ri} shows colour maps of the $\textrm{SoC}$ distribution on both tanks at $t/t_\textrm{end}=0.127$, the time evolution of the SoC of electrolyte $j$ at the half cell outlet, $\textrm{SoC}_\textrm{HC}^j$, and the mixing index of tank $j$, $\rm M^j$. As can be seen in Figure \ref{figure_Ri}, positive (negative) Richardson numbers prevail in the positive (negative) tank for the selected operating conditions, thus leading to positively (negatively) buoyant flows with well differentiated behaviors.

Solution A, with $|\textrm{Ri}^j| \gg 1$ in both tanks, is completely dominated by buoyancy. A strong stratification is produced in the negative side ($\rm Ri^- < 0$), causing a vertical piston flow that slowly fills the tank by pushing a clearly defined SoC interface between a lower region with $\rm SoC = SoC_0 = 0.01$ and an upper region with $\rm SoC = SoC_0 + \Delta SoC = 0.4131$, with $\Delta \rm SoC = 0.4031$ given for a single cell pass by equation \eqref{eq_DeltaSoC}. This solution is highly preferable as it ensures that the cell consistently receives fully discharged electrolyte until the interface reaches the tank's bottom, which occurs at $t/t_{\rm end} \simeq t_{\rm tank + pipe}/t_{\rm end} = \Delta {\rm SoC}$, when the first pass finishes and a new interface appears at the top of the tank. The piston flow repeats once and again for the following cell passes leading to a step-wise growth of $\textrm{SoC}_\textrm{HC}^-$ with constant (i.e., flat) periods, separated by abrupt jumps of magnitude $\Delta \rm SoC$ at regular intervals $\Delta t/t_{\rm end} \simeq \Delta \rm SoC$. On the other hand, the positive tank ($\rm Ri^+ > 0$) exhibits a slow deposition of the denser renewed electrolyte as small drops that fall directly to the tank bottom.

In both cases, the mixing index starts decreasing and eventually increases during the charge. In the positive tank, the portion of renewed electrolyte remained in the tank flow slowly, increasing the mixing index by about $0.1$ from its minimum value of ca.~0.74 up to $0.91$ at the end of the charge. However, in the negative tank it drops to lower values, reaching a minimum of M$^-=0.453$ as the fluid separates into two segregated regions. Subsequently, the mixing index rises again, peaking at about $0.87$ when the renewed electrolyte completely fills the tank. The evolution of $\textrm{SoC}_\textrm{HC}^j$ is very different in both half cells. The negative cell shows a plateau for $0 \lesssim t/t_{\rm end} \lesssim \Delta \rm SoC$ while the electrolyte with $\rm SoC_\textrm{HC}^- = SoC_0 + \Delta {\rm SoC}$ is being feed to the tank, a level that remains significantly lower than that predicted by the CSTR model. By contrast, the positive tank shows a highly unstable behaviour, corresponding to the intermittent drop-like behavior of the renewed electrolyte, which flows directly to the bottom and feeds the positive half-cell with electrolyte with higher SoC than predicted by the CSTR model. The difference between the $\textrm{SoC}_\textrm{HC}^j$ of the two half cells is as high as $0.414$ at $t/t_\textrm{end}=0.369$, which highlights the relevance of the fluid dynamics in determining cell performance under extreme conditions. Solution A stops earlier than Solutions B and C because its cut-off potential is reached sooner. This is because the charged electrolyte is directly deposited onto the bottom of the tank, which makes it appear fully charged from the perspective of the cell earlier than in the other cases.

Solution B is affected by both momentum and buoyancy, with momentum (buoyancy) being relatively more important in the positive (negative) tank according to the values of the Richardson numbers. This solution is qualitatively similar to solution A in the positive tank, but much closer to the CSTR model in the negative tank. The latter exhibits an initial piston-like flow with tiny jumps of $\Delta \rm SoC = 0.0242$ in which the electrolyte is renewed at a much faster rate than in the previous solution, now in a characteristic time of order $\Delta t/t_{\rm end} \simeq \Delta \rm SoC = 0.0242$. After a few cycles---slightly over 5 in the SoC$^-$ map of Figure \ref{figure_impact_Ri}---the combined effect of convection and diffusion homogenizes the electrolyte composition and the solution converges to a well-mixed nearly-homogeneous flow. The initial step-wise growth of $\textrm{SoC}_\textrm{HC}^-$ is thus smoothed out, eventually approaching the CSTR model, while $\textrm{M}^-$ reaches a very high value of ca.~$0.95$. In the positive tank, the positively buoyant jet tends to initially adhere to the left wall and subsequently to the bottom of the tank, until it reaches the outlet. This jet is unstable, producing several spikes in the $\textrm{SoC}_\textrm{HC}^+$ curve throughout the charge process and increasing the mixing index by about $0.1$, in a similar way as solution A, from its minimum value of ca.~0.775, after the initial transient, up to 0.88 at half charge. The values of $\textrm{SoC}_\textrm{HC}^+$ are still higher than those of the CSTR model, like in solution A, but the difference is smaller, with a very stable overshoot of $0.095$ that represents also the difference in $\textrm{SoC}_\textrm{HC}^j$ between both half cells.

Solution C is dominated by the momentum of the discharging jet, which establishes a quasi-steady flow with the jet crossing the tank directly from inlet to outlet. This flow pattern increases $\textrm{SoC}_\textrm{HC}^j$ and decreases M$^j$ with respect to the CSTR model, as only a small fraction of the renewed electrolyte recirculates and mixes with the fluid already in the tank. The secondary recirculating vortices bear a resemblance to those found in the rectangular lid-driven cavity flow and play a crucial role in the long-term homogenization of the electrolyte composition by diffusion. It is interesting to note that the negative tank exhibits slightly higher values of M$^-$ due to the additional corner eddy that appears at the top right of the tank, associated with the tendency of buoyancy to push the lighter fluid upwards in this region of low velocities. Anyway, the difference between both mixing indexes remains below $0.073$. The maximum difference between $\textrm{SoC}_\textrm{HC}^+$ and $\textrm{SoC}_\textrm{HC}^-$ is even smaller, $0.018$, much lower than in solutions A and B, and is reached at $t/t_\textrm{end}=0.360$.
After this point, the initial quasi-steady flows established in the tanks undergo a sudden destabilization due to transient buoyancy effects that leads to significant jet meandering. The resulting unsteady flow, established nearly simultaneously in both tanks, significantly increases mixing and approximates the solution to the CSTR model.

In all cases, the values of M$^j$ and $\textrm{SoC}_\textrm{HC}^j$ are closely related. The closer M$^j$ gets to unity, the more $\textrm{SoC}_\textrm{HC}^j$ resembles the CSTR model, and the more homogeneously the electrolyte fills the tank; whereas the smaller M$^j$, the more $\textrm{SoC}_\textrm{HC}^j$ differs from the CSTR model. In buoyancy dominated flows, $|\textrm{Ri}^j| \gg 1$, the values of $\textrm{SoC}_\textrm{HC}^j$ are larger than those of the CSTR model for $\textrm{Ri}^j > 0$ (denser electrolyte deposition to the bottom surface) and much lower for $\textrm{Ri}^j < 0$ (stratified piston flow). In momentum dominated flows, $|\textrm{Ri}^j| \ll 1$, the flow remains quasi-steady for a long induction period, with M$^j$ below and $\textrm{SoC}_\textrm{HC}^j$ above the CSTR solution. However, transient buoyancy effects eventually trigger a meandering jet instability in both tanks, which alters the nature of the flow and makes it highly unsteady, thus approaching the perfect mixing limit.

\subsection{Transient buoyancy effects} \label{section_transient}

The operating conditions studied above correspond to extreme values of the Richardson number, between which blends of different fluid behaviors may exist. As previously discussed, flows with $|\textrm{Ri}^j| \gg 1$ are dominated by buoyancy but exhibit markedly different behaviors for positive or negative Ri. In contrast, flows with $|\textrm{Ri}^j| \ll 1$ are initially dominated by momentum, but are also affected by buoyancy in the long term, with transient buoyancy effects starting earlier for larger Richardson numbers. Order-of-magnitude estimates cannot be used to analyze these subtle details, which must be studied numerically. This section presents two charge cycles for two values of $Q$, and hence two Richardson numbers, intermediate between those of solutions A, B and C.

\begin{figure}[ht!]
	\centering
	\includegraphics{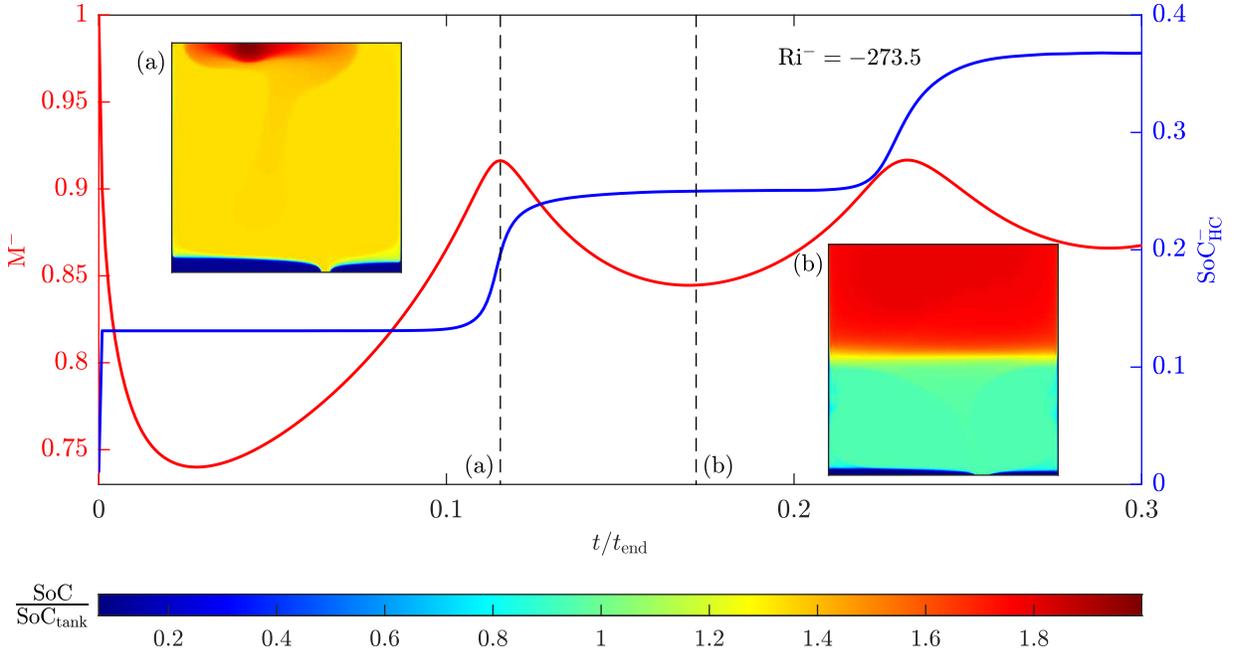}
	\caption{Time evolution of M$^-$ (red line, left axis) and $\textrm{SoC}_\textrm{HC}^-$ (blue line, right axis) corresponding to $Q = 1$ ml min$^{-1}$, showing two \textrm{SoC}$^-$ maps at dimensionless times (a) $0.116$ and (b) $0.172$. \label{figure_transient_buoyancy_Q1}}
\end{figure}

Figure \ref{figure_transient_buoyancy_Q1} shows the time evolution of M$^-$ and SoC$_{\rm HC}^-$ along a charge cycle with $Q = 1$ ml min$^{-1}$ (between solutions A and B) and two \textrm{SoC}$^-$ maps at two relevant times. This solution concerns values of $|\textrm{Ri}^j| \gg 1$ in both tanks (i.e., buoyancy dominated flows) but with different signs and a difference of one order of magnitude, $\textrm{Ri}^+ = 107.5$ and $\textrm{Ri}^- = -273.5$. As can be seen, the negative electrolyte suffers a strong stratification that results in a slow piston flow that fills the tank cyclically during the charge. This results in a step-wise growth of $\textrm{SoC}_\textrm{HC}^-$ that takes a time $\Delta t/t_{\rm end} \simeq \Delta \rm SoC = 0.121$ to fully renew the electrolyte in the tank. Due to the dominance of buoyancy over diffusion, there is no significant mixing of the electrolyte during the charging process of the negative tank, in contrast to solution B.

After a short initial transient, in which M$^-$ drops quickly from unity to a minimum of about $1-\Delta \rm SoC^{1/2}$ in times of order $t/t_{\rm end} \sim \rm SoC^{1/2}$, the mixing index increases again as the renewed electrolyte continues to fill the tank. When the tank is almost completely filled with partially charged electrolyte (with ${\rm SoC}^- = {\rm SoC_0} + \Delta \rm SoC$, see Figure \ref{figure_transient_buoyancy_Q1}a), M$^-$ reaches a maximum and $\textrm{SoC}_\textrm{HC}^-$ experiences a step-wise growth from ${\rm SoC_0} + \Delta \rm SoC$ to ${\rm SoC_0} + 2\Delta \rm SoC$. The value of M$^-$ does not reach unity because there is a thin layer of uncharged electrolyte that remains attached to the tank bottom until it is eventually drained through the outlet port at later times. As the partially charged electrolyte coming from the cell fills the tank with higher and higher $\textrm{SoC}_\textrm{HC}^-$, the cycle repeats once and again. M$^-$ repeatedly decreases to a minimum when the tank is half-filled (Figure \ref{figure_transient_buoyancy_Q1}h) and then rises back to a maximum, slightly higher than the previous one, when the tank is fully filled with renewed electrolyte. The step wise growth of $\textrm{SoC}_\textrm{HC}^-$ and the cyclic behaviour of M$^-$ are both dampened by the effect of diffusion and by the fact that as the electrolyte is charged the relative differences in SoC$^-$ throughout the tank become smaller, thus leading to higher values of M$^-$. As a result, the amplitude of the M$^-$ waves decreases along the charge process. The positive tank does not deserve special attention in this case, as its behavior is very similar to that exhibited by solutions A and B; the Richardson number is also positive and of order unity and M$^-$ reaches maximum values of about 0.85.
\begin{figure}[ht!]
	\centering
	\includegraphics{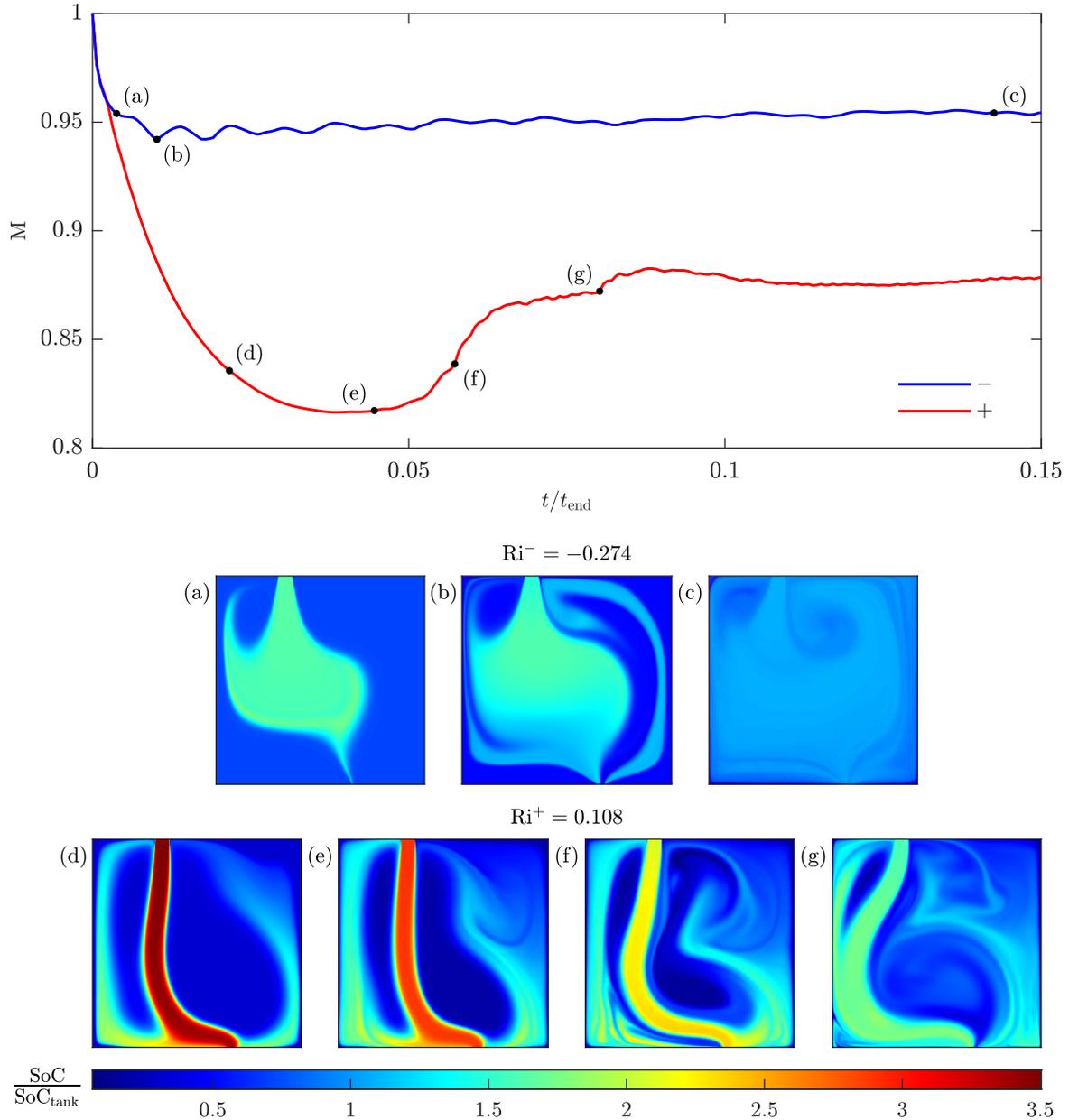}
	\caption{Time evolution of M using $Q = 10$ ml min$^{-1}$, showing three \textrm{SoC} maps corresponding to the negative tank at dimensionless times $3.82 \cdot 10^{-3}$ (a) $1.02 \cdot 10^{-2}$ (b) $0.143$ (c), and four \textrm{SoC} maps corresponding to the positive tank at $2.16 \cdot 10^{-2}$ (d) $4.46 \cdot 10^{-2}$ (e) $5.73 \cdot 10^{-2}$ (f) $8.02 \cdot 10^{-2}$ (g). \label{figure_transient_buoyancy_Q10}}
\end{figure}

Figure \ref{figure_transient_buoyancy_Q10} shows the time evolution of M$^-$ and M$^+$ during a charge cycle with a tenfold increase in the flow rate to $Q=10$~ml~min$^{-1}$, along with several SoC$^j$ maps for both tanks at relevant times. The solution exhibits unique fluid dynamics in each tank, with jet momentum playing a more significant role in both cases. In the negative tank, the effects of momentum and buoyancy are comparable, $\textrm{Ri}^-=-0.274$. The negative electrolyte (blue line) discharges as a negatively buoyant jet (Figure \ref{figure_transient_buoyancy_Q10}a), which quickly transforms into a plume that stays in the upper region of the tank, as illustrated in Figure \ref{figure_transient_buoyancy_Q10}b. The mixing index M$^-$ remains close to unity at all times, indicating an almost homogeneous electrolyte composition (Figure \ref{figure_transient_buoyancy_Q10}c), which closely approximates the CSTR model. By contrast, in the positive tank, $\textrm{Ri}^+=0.108$, and the flow is initially dominated by the  momentum of the positively buoyant jet. A quasi-steady flow is thus established at the initial times, as shown in Figure \ref{figure_transient_buoyancy_Q10}d. However, as the charge process proceeds the density differences within the tank become more relevant and buoyancy start to play a role, causing the main jet to eventually destabilize (Figures \ref{figure_transient_buoyancy_Q10}e-f). The secondary recirculation vortices interact with the main jet, leading to an averaging of the SoC$^+$ across the entire tank and a consistent increase in the mixing index M$^+$. This results in a more uniform electrolyte composition, as shown in Figure \ref{figure_transient_buoyancy_Q10}g at $t/t_\textrm{end}=8.02 \cdot 10^{-2}$, with the main jet attached to the left wall of the tank and the main recirculation region oscillating over time. The unsteady advective flows generated during the charging process homogenize species concentrations, leading to a quasi-steady increase in the mixing index M$^+$ through the last part of the charge.

The results of this section illustrate the variety of electrolyte flow regimes that emerge for less extreme values of Ri$^j$ than in the previous section. These behaviors are correctly anticipated by order of magnitude estimates for $1 \lesssim |\textrm{Ri}^j| \lesssim 10^3$, but the details can only be unveiled by numerical simulations. For the lower flow rate, $Q=1$~ml~min$^{-1}$, a stratified piston flow is established in the negative tank throughout the entire charge process, which coexists with a totally different flow in the opposite tank, where unstable plumes and jets produce instabilities from the beginning of the charge process, thereby homogenizing the flow.
At a higher flow rate of $Q=10$~ml~min$^{-1}$, with $\textrm{Ri}^+ \ll 1$, a quasi-steady jet is formed initially in the positive tank. However, the accumulation of density gradients in the recirculation eddies eventually disturbs the jet. The resulting unsteady flow leads to better-than-expected performance, with transient buoyancy slowly increasing the mixing index M$^+$ to values near to unity.

\subsection{Capacity loss} \label{section_capacity}

The above results show how different operation conditions may lead to different flow patterns within the tanks, producing $\textrm{SoC}_\textrm{HC}^j$ evolutions that may deviate to a large extent from the CSTR model and exhibit significant asymmetries. These asymmetries are not caused by electrolyte cross contamination due to the crossover of water and active species, which are ignored in this model, but to the mere imperfect operation of the tanks. Such non-ideal scenarios have a large impact on battery operation, causing the cell potential to significantly deviate from the symmetric perfect mixing case. This section explores these effects and investigates how fluid dynamics alone can modify the overall capacity and battery efficiency. This is achieved by simulating four charge/discharge cycles for solutions A, B, and C. To enable the study of each process separately, it is assumed that there is sufficient time between each charge and discharge for the electrolyte in the tanks to homogenize completely.

\begin{figure}[ht!]
	\centering
	\includegraphics{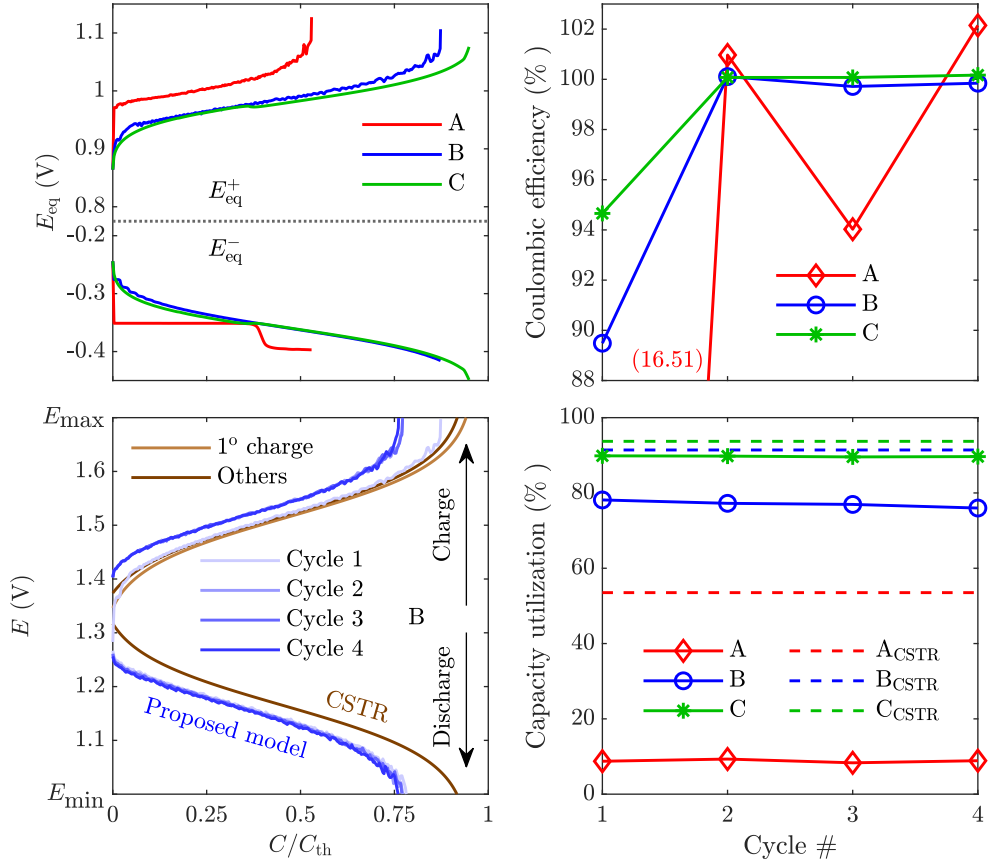}
	\caption{$E_\textrm{eq}^+$ and $E_\textrm{eq}^-$ for solutions A, B, and C during the first charge (top left), $E$ for solution B and the reference CSTR solution for 4 charge/discharge cycles (bottom left), coulombic efficiency (top right) and capacity utilization (bottom right) of the battery for solutions A, B, and C with realistic fluid mechanics and the CSTR model. \label{figure_capacity_loss}}
\end{figure}

Figure \ref{figure_capacity_loss} shows the equilibrium potential for both electrodes in solutions A, B, and C during the first cycle (top left), the cell potential for solution B and the reference CSTR solution for 4 charge/discharge cycles (bottom left), the coulombic efficiency for solutions A, B, and C (top right), and the capacity utilization (bottom right) of the battery for solutions A, B, and C under realistic fluid dynamics and using the CSTR model.

The equilibrium potentials show a clear impact of the flow pattern and the corresponding $\textrm{SoC}_\textrm{HC}^j$ evolution (see Figure \ref{figure_impact_Ri}) on battery performance. In solution A, the fast deposition of the charged electrolyte to the bottom in the positive tank increases $E_\textrm{eq}^+$ faster than expected. Simultaneously, the negative electrolyte exhibits a piston flow characterized by sudden jumps in $E_\textrm{eq}^-$. Both facts contribute to reaching the cut-off potential earlier than expected in the CSTR solution, thus decreasing the capacity utilization of the battery. Solutions B and C show similar performance in the negative electrolyte, with solution B giving slightly lower potentials (in absolute value) at the beginning (due to the piston flow) and solution C at the end (due to the jet instabilities). But the cell stops earlier in solution B due to the faster deposition of the positive electrolyte to the tank bottom.

In solution B (bottom left), the cell potential $E$ closely follows the CSTR solution during the first charge, but it diverges in the final phase when it starts to grow faster. The capacity loss after the first charge is caused, in particular, by the low mixing efficiency of the positive tank, which retains a certain amount of unmixed electrolyte by the end of the charge. Upon discharge, the flow patterns of the positive and negative tanks are reversed (due to the change in sign of $I$ and the Richardson number), resulting in the same overall behavior in the following cycles, which do not exhibit additional capacity losses. Thus, as the cycles continue, larger (smaller) cell potentials are observed during charge (discharge), causing $E$ to reach its upper (lower) threshold, $E_{\rm max}$ ($E_{\rm min}$), earlier than predicted by the CSTR model. This results in a consistent limited capacity utilization of about 76\% of the theoretical maximum in all subsequent cycles.

The coulombic efficiency (top right) remains consistently close to unity in all cases, increasing form a slightly lower value in the first cycle. This is because during the first cycle the electrolyte charges from an initial state of charge $\textrm{SoC}_0 \ll 1$. But the discharge stops once the cell potential reaches $E_\textrm{min}$, giving a final SoC that is consistently higher than $\textrm{SoC}_0$. All subsequent cycles, which start from this higher state of charge, reach coulombic efficiencies of roughly $100\%$, particularly for solutions B and C. As discussed below, solution A exhibits extremely low capacity utilization, which makes its coulombic efficiency also very low in the first charge-discharge cycle and more volatile in subsequent cycles.

The capacity utilization (bottom right) decreases with respect to the CSTR model in all cases. Solution A shows an extremely poor behavior resulting from the high concentration jump $\Delta \textrm{SoC}$ imposed by the cell at such low volumetric flow rates. In this case, the significantly denser drops of reacted electrolyte quickly accumulate at the bottom of the positive or negative tank in charge or discharge, bypassing most of the unreacted electrolyte that will never circulate through the cell. As a result, the cell potential reaches its cut-off values much earlier than in the CSTR model. This behavior persists over time, and even though the coulombic efficiency may seem high, this is only due to the poor performance of the battery in both charge and discharge. In fact, the capacity utilization remains below $10\%$, corresponding to an extremely undesirable scenario. In solution B, the fluid dynamics gets also reversed in charge and discharge, exhibiting an unstable jet that flows directly to the outlet in at least one of the tanks. As a result, the capacity utilization remains still moderately low, at roughly $76\%$, approximately $15\%$ below the CSTR model. By contrast, solution C exhibits quasi-steady flows during the first third of the charge, followed by a simultaneous destabilization of the main jets that start to meander between unstable recirculating regions. The highly homogeneous flow that results leads to a high capacity utilization of about $89.5\%$ during all cycles, which is very close to the $93.7\%$ predicted by the CSTR model. As a final remark, it is important to note that the capacity loss associated with imperfect mixing in solutions A and B might be worsened in 3D tanks, where the scale ratio between the volumetric flow rate and the entire volume of the tanks will be larger than in our 2D simulations.

\begin{table}[ht!]
	\centering
	\caption{Richardson numbers of the positive electrolyte for $Q = 5$~ml ~min$^{-1}$ (solution B) and the three values of $\rho_\textrm{SoC}^+$ under consideration.\label{table_rho_pos_Ri}}
	\begin{tabular}{rr}
		\hline
		\multicolumn{1}{c}{$\rho_\textrm{SoC}^+$} & \multicolumn{1}{c}{\multirow{2}{*}{$\textrm{Ri}^+$}} \rule{0pt}{2.6ex}\\
		  \multicolumn{1}{c}{(kg m$^{-3}$)} & \\
		\hline
		$-30$ & $-2.443\phantom{^*}$\\
		$10$ & $0.860^*$\\
        $30$ & $2.512\phantom{^*}$\\
		\hline
        \multicolumn{2}{c}{$^*$ baseline case}
	\end{tabular}
\end{table}

\subsection{$\rho_{\rm SoC}^+$ sensitivity analysis} \label{section_sensivity}

\begin{figure}[t!]
	\centering
	\includegraphics{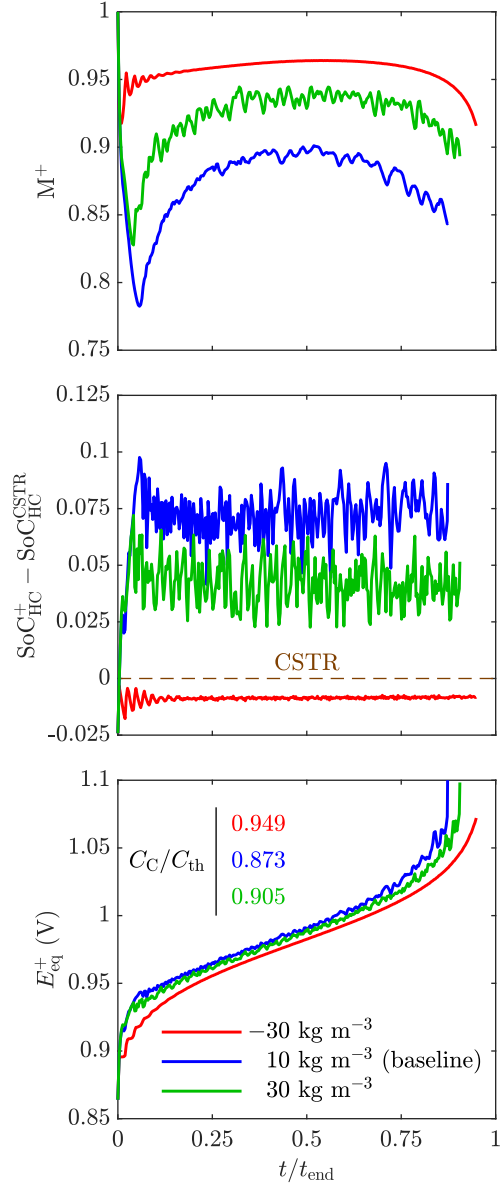}
	\caption{Positive mixing index (top), deviation of the state of charge of the positive half-cell with respect to the CSTR solution (medium) and equilibrium potential of the positive electrode (bottom) during charge corresponding to solution B and the three values of $\rho_\textrm{SoC}^+$ indicated in the figure. \label{figure_rho_pos}}
\end{figure}

The simulations presented above involve values of $\rho_\textrm{SoC}^-$ that are consistent with those reported in the literature (\ref{section_app_data}). However, a similar consensus on the value of $\rho_\textrm{SoC}^+$ has not yet been reached. This section presents a sensitivity analysis to asses the effect of this coefficient on the fluid dynamics of the positive tank as well as its impact on the equilibrium potential of the positive electrode and, therefore, the variation of the capacity utilization. Three simulations with $Q = 5$~ml ~min$^{-1}$ (solution B) were carried out with varying values of $\rho_\textrm{SoC}^+$ from $-30$ to $+30$ kg m$^{-3}$, including the baseline case of $10$ kg m$^{-3}$ used so far. The values lead to both positive and negative Richardson numbers, as listed in Table 4, which resulted in distinct electrolyte dynamics. It should be noted that while the three Richardson numbers are of order unity, cases with $\textrm{Ri}^+<0$ are expected to produce piston-like flows with plumes, while those with $\textrm{Ri}^+>0$ give unstable submerged jets and fast deposition of the renewed electrolyte to the tank bottom.

Figure \ref{figure_rho_pos} summarizes the results obtained in the simulations. Simulations conducted with $\rho_\textrm{SoC}^+=-30$ kg m$^{-3}$ (red line) exhibit the same behavior as the negative electrolyte (with $\rho_\textrm{SoC}^-=-30$ kg m$^{-3}$) so that a piston flows prevails also in the positive tank with the advantages mentioned above. This particular scenario produces the highest mixing index $\rm M^+$, which, in combination with the excellent performance of the negative side, leads to the highest capacity utilization. By contrast, the case $\rho_\textrm{SoC}^+=30$ (green line) exhibits a qualitatively similar behavior to the baseline case $\rho_\textrm{SoC}^+=10$ (blue line), with an unstable jet that crosses the tank and initially reduces mixing. However, as the Richardson number is larger than the baseline case, transient buoyancy effects appear earlier, thus increasing the mixing index $\rm M^+$ again after a short induction period with a net positive impact on battery capacity.

These results illustrate how minor variations in an unknown parameter can result in significant differences in battery performance. The impact of these variations can be further magnified by decreasing the flow rate or increasing the applied current. Therefore, accurate measurements of density and other physical properties of vanadium electrolytes are still needed to accurately predict the capacity utilization of this type of batteries.

\section{Conclusions}
\label{section_conclusions}

This work presents a detailed study of the fluid dynamics of the electrolytes in the tanks of vanadium redox flow batteries. The aim of the study is to investigate the effect of the fluid motion on the mixing of the electrolytes in the tanks, and its impact on battery performance and capacity utilization. The type of flow that emerges is determined by the competition between momentum and buoyancy in the jets that discharge into the tanks, with buoyancy forces being associated with the density changes that take place as the electrolytes pass through the cell. The overall features of the flow may differ significantly from the case of perfect mixing and can be roughly anticipated by a precise evaluation of the Richardson number, Ri. This was later confirmed by transient two-dimensional simulations based on the finite element software COMSOL Multiphysics. The numerical results show how the mixing index, the state of charge flowing to and from the cell, and the cell potential are all coupled, resulting in significant variations in overall energy capacity and system efficiency compared to the continuous stirred tank reactor model widely accepted in the literature. In particular, three different scenarios can be identified during charge based on the absolute value of the Richardson number, each one with different flow behaviors for positively, $\textrm{Ri}^j > 0$, and negatively, $\textrm{Ri}^j < 0$, buoyant jets:
\begin{itemize}

    \item $|\textrm{Ri}^j| \gg 1$: buoyancy is the dominant force, leading to a fast deposition of the renewed positive electrolyte ($\textrm{Ri}^+ > 0$) to the tank bottom, while the negative tank remains strongly stratified ($\textrm{Ri}^- < 0$), with the negative electrolyte filling the tank as a piston flows. The deposition flow in the positive tank reduces the battery capacity since $\textrm{SoC}^+_\textrm{HC}$ grows faster than expected, and thus, so does the cell potential.

    \item $|\textrm{Ri}^j| \sim 1$: inertia and buoyancy forces are comparable, leading to a highly unstable jet in the positive tank ($\textrm{Ri}^+ > 0$) that crosses the tank straight to the outlet thus decreasing the capacity. The negative electrolyte is initially stratified ($\textrm{Ri}^- < 0$), but due to the significant momentum of the jet the flow becomes more homogeneous, approaching perfect mixing.

    \item $|\textrm{Ri}^j| \ll 1$: buoyancy is negligible, and both tanks exhibit similar flow patterns, with two quasi-steady high-momentum jets that cross initially the tanks from inlet to outlet, inducing large recirculating vortices like in the lid-driven cavity flow. At large times, the jets are eventually affected by transient buoyancy due to the buildup of density variations in the recirculating bubbles, which recovers the mixing efficiency and has a positive effect in battery capacity.

\end{itemize}

Beyond these scenarios, more complex fluid dynamics may arise, such as plumes or transitions between different regimes. Thus, the highest coulombic efficiency is achieved for sufficiently low Richardson numbers for which the momentum of the incoming jet sets all the fluid in a stable recirculating motion in a first stage, but large enough to later induce transient buoyancy effects that eventually destabilize the main jet, thus increasing the mixing index and capacity utilization.

Our findings highlight the importance of having accurate values for the physical properties of vanadium electrolytes. In particular, additional experimental data is needed to consolidate the dependence of the electrolyte density with the state of charge, particularly for the positive electrolyte. The model should also be validated experimentally to determine its precision and contrast its results. Additionally, the current two-dimensional laminar flow model could serve for the development of a more advanced model adapted to industrial conditions with use made of turbulence models and considering more realistic geometries.

\section*{Acknowledgments}
This work has been partially funded by Grant IND2019/AMB-17273 of the Comunidad de Madrid, and by the Spanish Agencia Estatal de Investigaci\'{o}n projects TED2021-129378B-C21, PID2019-106740RB-I00, and RTC-2017-5955-3/AEI/10.13039/501100011033. Pablo A. Prieto-D\'iaz also acknowledges Ange A. Maurice for his critical feedback.

\section*{CRediT author statement}
\textbf{Pablo A. Prieto-D\'iaz}: Conceptualization, Methodology, Software, Validation, Formal Analysis, Investigation, Data Curation, Writing - Original Draft, Visualization. \textbf{Santiago E. Ib\'a\~{n}ez}: Methodology, Investigation, Writing - Review \& Editing. \textbf{Marcos Vera}: Conceptualization, Validation, Formal analysis, Investigation, Writing - Review \& Editing, Visualization, Resources, Supervision, Project administration, Funding acquisition.

\appendix
\section{The continuous stirred tank reactor model (CSTR)}\label{section_app_CSTR}

This appendix reviews the zero-dimensional Continuous Stirred Tank Reactor (CSTR) model. This model uses a simple ordinary differential equation to describe the time-delay response of the tank. As crossover and cross-contamination effects are ignored, the model can be written in terms of the $\textrm{SoC}$ instead of the concentration of the active vanadium species, and takes the same form for both tanks and both half-cells. In the CSTR model, the time evolution of the $\textrm{SoC}$ in the half-cells is described by the following integral conservation equation
\begin{equation}
    V_\textrm{HC}\frac{d\textrm{SoC}_\textrm{HC}}{dt} = Q\left( \textrm{SoC}_\textrm{tank} - \textrm{SoC}_\textrm{HC} \right) + \frac{I}{Fc_\textrm{tot}} \label{eq_PMM_c1}
\end{equation}
where the first term in the right hand side represents the effect of convective transport and the second that of electrochemical conversion, while that in the tanks is given by
\begin{equation}
    V_\textrm{tank}\frac{d\textrm{SoC}_\textrm{tank}}{dt} = Q\left( \textrm{SoC}_\textrm{HC} - \textrm{SoC}_\textrm{tank} \right) \label{eq_PMM_c2}
\end{equation}
where the conversion term is now absent. In the above equations $V_\textrm{HC}$ denotes the volume of the half-cells and $V_\textrm{tank}$ the volume of the tanks.

The analytical solution of these equations is
\begin{align}
    \textrm{SoC}_\textrm{HC}^\textrm{CSTR} &= \textrm{SoC}_0 + \frac{I}{Fc_\textrm{tot} V_\textrm{tank}\left(1+\epsilon \right)} \left\lbrace t - \frac{V_\textrm{tank}}{\left(1+\epsilon\right)Q} \left[ e^{-(1+\epsilon)\frac{Qt}{V_\textrm{HC}} } - 1 \right] \right\rbrace \label{eq_PMM_solc1} \\
    \textrm{SoC}_\textrm{tank}^\textrm{CSTR} &= \textrm{SoC}_0 + \frac{I}{Fc_\textrm{tot} V_\textrm{tank}\left(1+\epsilon \right)} \left\lbrace t + \frac{\epsilon V_\textrm{tank}}{\left(1+\epsilon\right)Q} \left[ e^{-(1+\epsilon)\frac{Qt}{V_\textrm{HC}} } - 1 \right] \right\rbrace \label{eq_PMM_solc2}
\end{align}
where $\epsilon = V_\textrm{HC}/V_\textrm{tank} \ll 1$ is the half-cell to tank volume ratio, usually small in applications, $\textrm{SoC}_0$ is the state of charge at the initial time, $t=0$, assumed to be equal for both tanks and half cells, and the applied current $I$ is considered constant over time, being positive during charge and negative in discharge. The exponential term decays in times of the order of the cell residence time, $V_\textrm{HC}/Q$. This time is of order $\epsilon$ compared to the residence time in the tanks, $t_{\rm tank} = V_\textrm{tank}/Q$, which is in turn small compared to the time required to fully charge and discharge the flow battery
\begin{equation}
    t_\textrm{end} \sim \frac{Fc_\textrm{tot}V_\textrm{tank}\left( \textrm{SoC}_\textrm{end}-\textrm{SoC}_0 \right)}{I} \label{eq_tend}
\end{equation}
estimated from \eqref{eq_PMM_solc2} with $\textrm{SoC}_\textrm{end}$ denoting the state of charge at $t_\textrm{end}$. In fact, the ratio of the residence time in the tanks to the discharge time can be anticipated to be
\begin{equation}
    \frac{t_{\rm tank}}{t_\textrm{end}} \sim \frac{I/(Fc_\textrm{tot}Q)}{\left( \textrm{SoC}_\textrm{end}-\textrm{SoC}_0 \right)} =  \frac{\Delta {\rm SoC}}{\left( \textrm{SoC}_\textrm{end}-\textrm{SoC}_0 \right)} \ll 1
\end{equation}
where $\Delta \textrm{SoC}$ denotes the change in SoC in a single cell pass given by~\eqref{eq_DeltaSoC}. This ratio is small compared to unity because $\Delta \textrm{SoC}$ is usually much smaller than unity, while the total SoC change during a full charge or discharge cycle is, by definition, of order unity. The SoC at the end of the process, $\textrm{SoC}_\textrm{end}$, can be computed using the cut-off potentials, $E_k = E_\textrm{max}$ or $E_\textrm{min}$ for charge or discharge, respectively, along with the polarization curve \eqref{eq_E_curve} to give
\begin{equation}
    \textrm{SoC}_\textrm{end} = {\left[ 1+e^{-\left( E_k-E_0-IR_\textrm{C} \right)F/2RT}  \right]}^{-1}.
\end{equation}

Under the above considerations, after the decay of the exponential term, and in the limit $\epsilon \rightarrow 0$, equations \eqref{eq_PMM_solc1} and \eqref{eq_PMM_solc2} can be rewritten solely in terms of $\rm SoC_0$, $\Delta\rm SoC$ and $t/t_{\rm tank}$ to give
\begin{align}
    \textrm{SoC}_\textrm{HC}^\textrm{CSTR} &= \textrm{SoC}_0 + \left(\frac{t}{t_{\rm tank}} + 1\right) \Delta \textrm{SoC} \label{eq_PMM_solfinalc1} \\
    \textrm{SoC}_\textrm{tank}^\textrm{CSTR} &= \textrm{SoC}_0 + \frac{t}{t_{\rm tank}} \Delta \textrm{SoC}  \label{eq_PMM_solfinalc2}
\end{align}
to be used in this paper as reference CSTR solution for comparative purposes.

\section{Density of vanadium electrolytes: A short literature review} \label{section_app_data}

This appendix reviews the available data on the density of vanadium electrolytes. Density is an important physical property of the electrolytes, but has not received as much attention as viscosity. In particular, there is a discrepancy in the level of agreement regarding the density variations with temperature and composition in the positive and negative electrolytes \cite{Skyllas2016, Mousa2003, Rahman2009, Xu2014-2}.

The density of the negative electrolyte has been extensively studied, and several authors have established a common value for $\rho_T^{-} = -0.6$ kg m$^{-3}$ K$^{-1}$ for different total vanadium concentrations \cite{Skyllas2016, Mousa2003, Ressel2018}. Tables \ref{table_density_Ressel} and \ref{table_density_Skyllas} gather the values of $\rho_\textrm{SoC}^-$ reported by Ressel et al.~\cite{Ressel2018} at 293 K for several cycles, and by Skyllas et al.~\cite{Skyllas2016} for three operating temperatures. Since these researchers used different total vanadium concentrations, $c_{\rm tot}$, the reference density $\rho_0^-$ was slightly different. However, the values of $\rho_\textrm{SoC}^-$ are very similar in all cases, so a constant value of $\rho_\textrm{SoC}^- = -30$ kg m$^{-3}$ was used in this work.

In contrast, there is limited data on the density variation of the positive electrolyte. Skyllas et al.~\cite{Skyllas2016} reported the values of $\rho_T^{+}$ and $\rho_\textrm{SoC}^+$ listed in Table \ref{table_density_Skyllas}. The data lead to an averaged value for $\rho_T^{+}$ of $-0.6$ kg m$^{-3}$ K$^{-1}$ for the various SoC$^+$ under study, and support the selection of the baseline value $\rho_\textrm{SoC}^+ = 10$ kg m$^{-3}$ in the main text. However, the absence of error bars in \cite{Skyllas2016} made it impossible to estimate the errors in the density gradients, which underlined the need for the sensitivity analysis presented in section~\ref{section_sensivity}.

Regarding modeling studies, they typically assume a constant density for both electrolytes, with values ranging from 1300 kg/m$^3$ \cite{Hao2019} or 1354 kg m$^{-3}$ \cite{Tang2012-2,Tang2012,Trovo2019-2,Trovo2019,Yan2016} to 1400 kg/m$^3$ \cite{Xiong2013, Zhou2015, We2014, Wei2014}. Few studies distinguish between the densities of the two electrolytes. For instance, Knehr et al.~\cite{Knehr2012} used a density of 1300 kg/m$^3$ for the negative electrolyte and 1350 kg/m$^3$ for the positive electrolyte. Notably, while these studies assumed similar densities, they used different total vanadium and sulfate concentrations, suggesting that the role of density is usually overlooked in VRFB models.

\begin{table}[ht!]
	\centering
	\caption{Values of $\rho_\textrm{SoC}^-$ at $293$ K reported by Ressel et al.~\cite{Ressel2018}.\label{table_density_Ressel}}
	\begin{tabular}{cc}
	    \hline
        \multirow{2}{*}{Cycle \#} & $\rho_\textrm{SoC}^-$ \rule{0pt}{2.6ex}\\
		  & (kg m$^{-3}$) \\
		\hline
        $1$ & $-30.29$\\
        $2$ & $-29.95$\\
        $3$ & $-29.56$\\
        $4$ & $-29.56$\\
        $5$ & $-29.29$\\
        $6$ & $-28.84$\\ \hline
		\end{tabular}
	\caption{Values of $\rho_\textrm{SoC}^-$ and $\rho_\textrm{SoC}^+$ at different temperatures obtained from the data reported by Skyllas et al.~\cite{Skyllas2016}.\label{table_density_Skyllas}}
	\begin{tabular}{ccc}
        \hline
		\multirow{2}{*}{$T$ (K)} & $\rho_\textrm{SoC}^-$ & $\rho_\textrm{SoC}^+$ \rule{0pt}{2.6ex}\\
		  & (kg m$^{-3}$) & (kg m$^{-3}$) \\
		\hline
		$283$ & $-27.43$ & $8.29$ \\
		$293$ & $-30.43$ & $9.00$ \\
		$303$ & $-29.71$ & $8.43$ \\ \hline
	\end{tabular}
\end{table}

\section*{Supplementary Material}
For supplementary material see the Video Abstract here:\\
{https://www.editorialmanager.com/hmt/download.aspx?id=976436\&guid=db6e6119-7350-4f0d-91cd-f2ab7efc0cbe\&scheme=1}.

\bibliographystyle{elsarticle-num}
\bibliography{refs.bib}


\setlength{\nomlabelwidth}{3cm}

\nomenclature{$(x, y)$}{Horizontal and vertical cartesian coordinates}
\nomenclature{$\vec{g}$}{Acceleration of gravity $(0,g)$ (m s$^{-2}$)}
\nomenclature{$L_\textrm{tank}$}{Tank side (m)}
\nomenclature{$D_{\textrm{P}}$}{Pipe diameter (m)}
\nomenclature[G]{$\Sigma$}{Surface (m$^2$)}
\nomenclature[S]{pipe}{Pipe}
\nomenclature[S]{tank}{Tank}
\nomenclature[S]{C}{Cell}
\nomenclature[S]{HC}{Half-cell}
\nomenclature{$t$}{Time (s)}
\nomenclature{$t_{\textrm{end}}$}{Time to completely charge/discharge (s)}
\nomenclature{$s$}{Standard deviation}
\nomenclature{CSTR}{Continuous Stirred Tank Reactor}
\nomenclature[S]{$0$}{Reference value}
\nomenclature[S]{in}{Tank inlet}
\nomenclature[S]{out}{Tank outlet}
\nomenclature[S]{$i$}{Vanadium species, $\rm \{II, III, IV, V\}$}
\nomenclature[P]{$j$}{Positive or negative electrolyte/tank, $\{+,-\}$}
\nomenclature{$T$}{Temperature (K)}
\nomenclature{$T_\textrm{amb}$}{Ambient temperature (K)}

\nomenclature{$\vec{u}$}{Velocity vector $(u, v)$ (m s$^{-1}$)}
\nomenclature{$(u, v)$}{Velocity components along the $(x,y)$-axes (m s$^{-1}$)}
\nomenclature{$Q$}{Volumetric flow rate (m$^3$ s$^{-1}$)}
\nomenclature{$q$}{Flow rate per unit length (m$^2$ s$^{-1}$)}
\nomenclature{$p$}{Pressure (Pa)}
\nomenclature[G]{$\mu$}{Viscosity (Pa s)}
\nomenclature[G]{$\rho$}{Density (kg m$^{-3}$)}
\nomenclature{$c_p$}{Specific heat at constant pressure (J kg$^{-1}$ K$^{-1}$)}
\nomenclature{$D_i$}{Diffusion coefficient of species $i$ (m$^2$ s$^{-1}$)}
\nomenclature{$k$}{Thermal conductivity (W m$^{-1}$ K$^{-1}$)}

\nomenclature{$\textrm{Re}$}{Reynolds number}
\nomenclature{$\textrm{Fr}$}{Froude number}
\nomenclature{$\textrm{Pe}$}{Peclet number}
\nomenclature{$\textrm{Ri}$}{Richardson number}
\nomenclature{M}{Mixing Index}

\nomenclature{$I$}{Total applied current (A)}
\nomenclature{$F$}{Faraday constant}
\nomenclature{$R_\textrm{C}$}{Ohmic resistance ($\Omega$)}
\nomenclature{$\dot{Q}_\textrm{J}$}{Joule heating per unit time (W)}
\nomenclature{$\dot{Q}_\textrm{R}$}{Heat due to electrochemical reactions per unit time (W)}
\nomenclature{$\Delta S$}{Entropy change (J mol $^{-1}$ K $^{-1}$)}
\nomenclature{$E$}{Cell potential (V)}
\nomenclature{$E_\textrm{eq}^j$}{Equilibrium potential for half-cell $j$ (V)}
\nomenclature{$E_0$}{Formal cell potential (V)}
\nomenclature{$C$}{Capacity (C)}

\nomenclature{$c$}{Concentration (M)}
\nomenclature{$\textrm{SoC}$}{State of Charge}
\nomenclature[P]{$-$}{Negative electrolyte/tank}
\nomenclature[P]{$+$}{Positive electrolyte/tank}
\nomenclature[S]{II}{$\textrm{V}^{2+}$}
\nomenclature[S]{III}{$\textrm{V}^{3+}$}
\nomenclature[S]{IV}{$\textrm{VO}^{2+}$}
\nomenclature[S]{V}{$\textrm{VO}_2^+$}
\nomenclature{$c_\textrm{tot}$}{Total vanadium concentration (M)}



\end{document}